\documentclass[aps,pra]{revtex4}
\usepackage{amsmath}
\usepackage{graphicx}
\usepackage{epsfig}
\usepackage{braket}
\usepackage{amsmath,amsfonts,amssymb}
\usepackage{xparse}
\usepackage{tabularx}
\usepackage{pbox}

\begin{document}

\title{Instantaneous measurements of nonlocal variables in relativistic quantum theory \\ (a review)}

\author{Matthew J. Lake${}^{a,b}$\footnote{matthewj@nu.ac.th}}
\affiliation{${}^b$ The Institute for Fundamental Study, ``The Tah Poe Academia Institute", \\ Naresuan University, Phitsanulok 65000, Thailand and \\ Thailand Center of Excellence in Physics, Ministry of Education, Bangkok 10400, Thailand}

\date{\today}

\begin{abstract}
This article reviews six historically important papers in the development of the theory of measurement for nonlocal variables in quantum mechanics, with special emphasis the non violation of relativistic causality. Spanning more than seventy years, we chart the major developments in the field from the declaration, by Landau and Peierls in 1931, that measurement of nonlocal variables was impossible in the relativistic regime to the demonstration, by Vaidman in 2003, that all such variables \emph{can} be measured instantaneously without violation of causality through an appropriate act of ``measurement", albeit not of a standard projective (Von Neumann) type. 
\end{abstract}
\maketitle
\tableofcontents

\section{Introduction}%``
\label{SectI}

The question of whether nonlocal variables could be measured instantaneously was considered by no means obvious in the early days of quantum theory. It was initially thought that, although measurements of local variables presented no problems, the principle of relativistic causality placed severe restrictions upon the measurability of nonlocal operators. 

This idea was first proposed by Landau and Peierls in 1931 \cite{LandauPeierls1931}. They claimed that the measurement of \emph{any} nonlocal observable necessarily violated causality and, therefore, indicated the failure of quantum mechanics in the relativistic range. Subsequent generations of physicists, who were less skeptical about the applicability of quantum theory to relativistic problems, concluded that the measurement of nonlocal variables must therefore be impossible. This was until 1981, when the first experiment designed specifically to measure a nonlocal property was proposed by Aharonov and Albert \cite{AharonovAlbert1981}.

This essay reviews six papers, including the ones mentioned above, written between 1931 and 2003, which explore questions relating to the instantaneous measurement of nonlocal variables in relativistic quantum theory. Such measurements must be possible if we are to grant nonlocal variables the status of observables. We conclude that all nonlocal operators \emph{may} be measured instantaneously and that, in this sense, the predictions of quantum mechanics are entirely consistent with the requirements of special relativity. However, as we shall demonstrate, the principle of relativistic causality \emph{does} place constraints upon the \emph{way} in which such measurements may be performed.

%%%%%%%%%%%%%%%%%%%%%%%%%%%%%%%%%%%%%%%%%%%%%%%%%%%%%%%
%%%%%%%%%%%%%%%%%%%%%%%%%%%%%%%%%%%%%%%%%%%%%%%%%%%%%%%
\section{Nonlocality in nonrelativistic quantum theory}%``
\label{SectII}

%%%%%%%%%%%%%%%%%%%%%%%%%%%%%%%%%%%%%%%%%%%%%%%%%%%%%%%
\subsection{Nonlocal states and nonlocal variables}%``
\label{SectII.1}

The principle of superposition in nonrelativistic quantum mechanics allows the existence of so called nonlocal states for composite systems. A useful example is a system composed of two fermions. For simplicity we may assume that each fermion is fixed at a point in space, either $x_A$ or $x_B$, for example, by some potential, and consider only the spin part of the total wave function. The general state space of a composite system is given by the tensor product of the Hilbert spaces which form the state spaces of the individual subsystems,
\begin{eqnarray} \label{subsys}
\prod_{i=1}^{N} \otimes H_{i},
\end{eqnarray}
where $N$ is the number of subsystems. In this case the appropriate state space is a four-dimensional Hilbert space which is the tensor product of the two, two-dimensional, Hilbert spaces of the individual particles. The space is spanned by any four mutually orthogonal state vectors, of which the four direct
product states, 
\begin{subequations}
\begin{align}
&\Ket{\Psi_1} = \Ket{\uparrow_z}_{A}  \otimes \Ket{\uparrow_z}_{B}, \ \  \Ket{\Psi_2} = \Ket{\downarrow_z}_{A}  \otimes \ket{\downarrow_z}_{B},
\label{basis1.a} \\
&\Ket{\Psi_3} = \Ket{\uparrow_z}_{A}  \otimes \Ket{\downarrow_z}_{B}, \ \  \Ket{\Psi_4} = \Ket{\downarrow_z}_{A}  \otimes \Ket{\uparrow_z}_{B}, 
\label{basis1.b} 
\end{align}
\end{subequations}
are an obvious example. However, any four mutually orthogonal states form an equally valid basis. The spectral theorem for self-adjoint operators \cite{Isham1995} states that an operator may be defined entirely in terms of its eigenvectors,
\begin{eqnarray} \label{spectraltheorem}
\hat{A} = \sum_{m=1}^{M} \sum_{j=1}^{d(m)}a_m \ket{a_m, j} \Bra{a_m, j},
\end{eqnarray}
where the eigenvalue $a_m$ has degeneracy $d(m)$, of which $\dim(H) =  \sum_{m=1}^{M}d(m)$ vectors will be linearly independent. It is therefore possible to use any set of $\dim(H)$ linearly independent eigenvectors of any Hermitian operator as a complete set of mutually orthogonal basis vectors. However, the eigenstates of a general Hermitian operator are degenerate and it is necessary to specify the simultaneous eigenvalues of two or more commuting operators in order to specify a state completely. For example, in our system the direct product states may be rewritten as the
simultaneous eigenvectors of the operators $\hat{\sigma}_z^{(A)}$ and $\hat{\sigma}_z^{(B)}$ which represent the $z$-component of spin for each of the one-particle subsystems,
\begin{subequations}
\begin{align}
&\Ket{\Psi_1} = \Ket{\uparrow_z}_{A}  \otimes \Ket{\uparrow_z}_{B} = \Ket{\hat{\sigma}_z^{(A)} = +\hbar/2, \hat{\sigma}_z^{(B)} = +\hbar/2}, 
\label{basis2.a} \\
&\Ket{\Psi_2} = \Ket{\downarrow_z}_{A}  \otimes \ket{\downarrow_z}_{B} = \Ket{\hat{\sigma}_z^{(A)} = -\hbar/2, \hat{\sigma}_z^{(B)} = -\hbar/2}, 
\label{basis2.b} \\
&\Ket{\Psi_3} = \Ket{\uparrow_z}_{A}  \otimes \Ket{\downarrow_z}_{B} = \Ket{\hat{\sigma}_z^{(A)} = +\hbar/2, \hat{\sigma}_z^{(B)} = -\hbar/2}, 
\label{basis2.c} \\
&\Ket{\Psi_4} = \Ket{\downarrow_z}_{A}  \otimes \Ket{\uparrow_z}_{B} = \Ket{\hat{\sigma}_z^{(A)} = -\hbar/2, \hat{\sigma}_z^{(B)} = +\hbar/2}.
\label{basis2.d} 
\end{align}
\end{subequations}
In addition to the four direct product states, in which each fermion has either spin $+\hbar/2$, denoted $\Ket{\uparrow_z}$, or spin $-\hbar/2$, denoted $\Ket{\downarrow_z}$, states, there are an infinite number of possible superpositions of these states. Four superpositions of particular interest are the famous Bell states \cite{Bell1964}, in which the coefficients of each product term are equal,
\begin{subequations}
\begin{align} 
&\Ket{\Psi_{\pm}} = \frac{1}{\sqrt{2}}\left(\Ket{\uparrow_z}_{A}  \otimes \Ket{\downarrow_z}_{B} \pm  \Ket{\downarrow_z}_{A}  \otimes \Ket{\uparrow_z}_{B}\right),
\label{Bellstates.a} \\
&\Ket{\Phi_{\pm}} =  \frac{1}{\sqrt{2}}\left(\Ket{\uparrow_z}_{A}  \otimes \Ket{\uparrow_z}_{B} \pm  \Ket{\downarrow_z}_{A}  \otimes \Ket{\downarrow_z}_{B}\right).
\label{Bellstates.b} 
\end{align}
\end{subequations}
In the two fermion system $\hat{\sigma}_z^{(A)}$ and $\hat{\sigma}_z^{(B)}$, which are represented by the appropriate Pauli spin matrix for single particles, are represented by $\hat{\sigma}_z^{(A)} \otimes \mathbb{I}_{B}$ and $\mathbb{I}_{A} \otimes \hat{\sigma}_z^{(B)}$, where $\mathbb{I}$ is the identity matrix. In a similar manner, the operators $\hat{\sigma}_z$ and $\hat{\sigma}^{2}$ can be defined as
\begin{eqnarray} \label{spinT}
\hat{\sigma}_z = \left(\hat{\sigma}_z^{(A)} \otimes \mathbb{I}_{B}\right) + \left(\mathbb{I}_{A} \otimes \hat{\sigma}_z^{(B)}\right),
\end{eqnarray}
\begin{eqnarray} \label{spinSqr}
\hat{\sigma}^{2} = \hat{\sigma}_A^{2} + \hat{\sigma}_B^{2},
\end{eqnarray}
where
\begin{subequations}
\begin{align}
\hat{\sigma}_A = \left(\hat{\sigma}_x^{(A)} + \hat{\sigma}_y^{(A)} + \hat{\sigma}_z^{(A)}\right) \otimes \mathbb{I}_{B}, \label{spinAspinB.a} \\ 
\hat{\sigma}_B = \mathbb{I}_{A} \otimes \left(\hat{\sigma}_x^{(B)} + \hat{\sigma}_y^{(B)} + \hat{\sigma}_z^{(B)}\right). \label{spinAspinB.b} 
\end{align}
\end{subequations}
These represent the $z$-component and squared magnitude of the total spin vector for the composite system. It can easily be verified that these operators commute and have eigenvectors corresponding to the eigenvalues $0$, $\pm \hbar$ and $0$, $+2\hbar$, respectively, in our example. The complete set of their simultaneous eigenvectors therefore forms an alternative basis for the four-dimensional Hilbert space, given by
\begin{subequations}
\begin{align}
&\Ket{\Psi_1} = \Ket{\uparrow_z}_{A}  \otimes \Ket{\uparrow_z}_{B} = \Ket{\hat{\sigma}_z = +\hbar, \hat{\sigma}^{2} = +2\hbar^2}, 
\label{eigenvecs1.a} \\ 
&\Ket{\Psi_2} = \Ket{\downarrow_z}_{A}  \otimes \ket{\downarrow_z}_{B} =  \Ket{\hat{\sigma}_z = -\hbar, \hat{\sigma}^{2} = +2\hbar^2}, \label{eigenvecs1.b} \\ 
&\Ket{\Psi_{+}} = \frac{1}{\sqrt{2}}\left(\Ket{\uparrow_z}_{A}  \otimes \Ket{\downarrow_z}_{B} +  \Ket{\downarrow_z}_{A}  \otimes \Ket{\uparrow_z}_{B}\right) = \Ket{\hat{\sigma}_z = 0, \hat{\sigma}^{2} = +2\hbar^2},
\label{eigenvecs1.c} \\ 
&\Ket{\Psi_{-}} = \frac{1}{\sqrt{2}}\left(\Ket{\uparrow_z}_{A}  \otimes \Ket{\downarrow_z}_{B} -  \Ket{\downarrow_z}_{A}  \otimes \Ket{\uparrow_z}_{B}\right) = \Ket{\hat{\sigma}_z = 0, \hat{\sigma}^{2} = 0}.
\label{eigenvecs1.d}
\end{align}
\end{subequations}
The key point to notice is that it is possible for the two particle system to possess a total spin vector with a squared magnitude of $0$ or $2\hbar^2$, and with zero $z$-component, when neither of its constituent particles can be said to possess a spin of $\pm \hbar/2$ individually.

\emph{Local variables}: For pure quantum states, a local state is an eigenstate of operators representing observables which the system can possess at a fixed point in space. For example, our second particle may possess a spin value of $\hat{\sigma}_z^{(B)} = \pm\hbar/2$ at $x_B$ . For the purposes of this essay, which deals only with pure states, such operators are called local operators. By this definition, $\hat{\sigma}_z^{(A)} \otimes \mathbb{I}_{B}$ and $\mathbb{I}_{A} \otimes \hat{\sigma}_z^{(B)}$ are local operators and $\Ket{\Psi_{1}}$, $\Ket{\Psi_{2}}$, $\Ket{\Psi_{3}}$ and $\Ket{\Psi_{4}}$, given by Eqs. (\ref{basis2.a})-(\ref{basis2.d}), are local states. We will see that all local operators may be measured instantaneously, and may therefore be referred to as local observables.

\emph{Nonlocal variables}: On the other hand, operators such as $\hat{\sigma}_z$ and a $\hat{\sigma}^{2}$ represent variables that are not spatially localised. These variables will be called nonlocal variables, or nonlocal observables if they can be measured in some way. These are
not strict definitions. In general, for mixed states, direct products exist which are nonlocal. This phenomenon is called nonlocality without entanglement \cite{NielsenChuang2000}, but will not be discussed here. As we shall see, although most, but not \emph{all}, local variables may be measured easily, the requirements of relativistic causality place certain constraints upon measurements of nonlocal variables.

Hence, by this definition, states such as $\Ket{\Psi_{+}}$ and $\Ket{\Psi_{-}}$, and other superpositions of direct product states, are referred to as nonlocal as they do not possess values of observables that exist at a specific point in space. Therefore, although local states may be the eigenvectors of both local and nonlocal operators, for example $\Ket{\Psi_{1}}$ and $\Ket{\Psi_{2}}$, nonlocal states may \emph{not} be the eigenstates of local operators. In our two-fermion system, both $\Ket{\Psi_{+}}$ and $\Ket{\Psi_{-}}$ possess values of the total spin vector squared \emph{and} its $z$-component, as they are the simultaneous eigenvectors of $\hat{\sigma}_z$ and $\hat{\sigma}^{2}$ but, unlike the states $\Ket{\Psi_{1}}$ to $\Ket{\Psi_{4}}$, neither particle possesses a spin individually at each spatial point, $x_A$ or $x_B$. The states $\Ket{\Psi_{+}}$ and $\Ket{\Psi_{-}}$ are therefore \emph{nonlocal}.

States of this kind are sometimes also referred to as entangled, as the probabilities of results of local measurements made on one subsystem are dependent upon the results of local measurements already made on another subsystem. Here, for example, beginning with either state $\Ket{\Psi_{\pm}}$ and measuring $\hat{\sigma}_z^{(A)} \otimes \mathbb{I}_{B}$ at point $x_A$, at time $t = t_0$, causes a collapse of the wave function and will yield the result $\pm \hbar/2$ with equal probability. The result of a subsequent measurement of $\mathbb{I}_{A} \otimes \hat{\sigma}_z^{(B)}$ at point $x_B$ at $t = t_0 + \epsilon$ will then, with certainty, yield $+\hbar/2$ . Thus the two sub-systems are unavoidably ``entangled" in some way. Again, this is not a strict definition, but the example is useful for our purposes, in order to illustrate the idea of entanglement. Formally, an entangled state is a state belonging to the Hilbert space of a composite system $\Ket{\Psi}_{AB} \in H_A \otimes H_B$ which may not be written as a direct product of states belonging to the Hilbert spaces of the individual subsystems, i.e. $\Ket{\Psi}_{AB} \neq \Ket{\Psi}_{A} \otimes \Ket{\Psi}_{B}$, where $\Ket{\Psi}_{A} \in H_A$, $\Ket{\Psi}_{B} \in H_B$. All such states belonging to any quantum system exhibit behaviour similar to that outlined above.

It is also worth noting here that nonlocal variables also exist for single particle states. The most common example is momentum. Momentum eigenstates are superpositions of infinitely many position eigenstates and are therefore nonlocal, although this point is often not emphasized in introductory courses on quantum mechanics.

\emph{In summary, the nonlocal states considered here are eigenvectors of only nonlocal operators, which represent quantities a system cannot posses at a single point in space.}

%%%%%%%%%%%%%%%%%%%%%%%%%%%%%%%%%%%%%%%%%%%%%%%%%%%%%%%%
\subsection{Measurement of nonlocal variables}%``
\label{SectII.2}

It was realised, very early on in the development of quantum mechanics that some superposition states possessed values of nonlocal variables, and that operators representing these variables could be mathematically constructed. It was also possible, to some extent, to measure the value of certain nonlocal variables. This could be done by using an ensemble of many copies of identically prepared systems and performing repeated local measurements. For example, in our system, if measurements of  $\hat{\sigma}_z^{(A)} \otimes \mathbb{I}_{B}$ and $\mathbb{I}_{A} \otimes \hat{\sigma}_z^{(B)}$ were performed on one thousand copies of identically prepared states and it was found that $\hat{\sigma}_z^{(A)} = +\hbar/2$ and $\hat{\sigma}_z^{(B)} = -\hbar/2$ approximately five hundred times, while $\hat{\sigma}_z^{(A)} = -\hbar/2$ and  $\hat{\sigma}_z^{(B)} = +\hbar/2$ approximately five hundred times, we would know that the original state was either $\Ket{\Psi}_{+}$ or $\Ket{\Psi}_{-}$, and so that the value of $\hat{\sigma}_z$ is zero. However, we would be unable to specify exactly which state was initially present, and hence unable to determine $\hat{\sigma}^{2}$.

This method suffers from three defects. Firstly, it is by no means obvious that it may be used to determine the value of a general nonlocal variable. Secondly, it is not instantaneous, by which we mean that the measurement cannot be completed at a specific instant in time and, thirdly, it destroys the state which we were initially interested in, though it may be recreated later using the same preparation which gave rise the the original ensemble. As stated in the Introduction, it was, in fact, highly uncertain until the early 1980's \cite{AharonovAlbert1981}, whether instantaneous measurements of \emph{any} nonlocal variables were physically possible.

%%%%%%%%%%%%%%%%%%%%%%%%%%%%%%%%%%%%%%%%%%%%%%%%%%%%%%%
%%%%%%%%%%%%%%%%%%%%%%%%%%%%%%%%%%%%%%%%%%%%%%%%%%%%%%%
\section{Can nonlocal variables be measured instantaneously in the relativistic range? ``No" - Landau and Peierls (1931)}%``
\label{SectIII}

Although it was commonly accepted that nonlocal variables existed, and the operators representing some such variables were well known, it was by no means clear in the early days of quantum theory whether any nonlocal property besides momentum $\hat{p}$, and operators of the form $F(\hat{p})$, could ever be measured instantaneously in the conventional Von Neumann \cite{VonNeumann1955} sense. In fact, as early as 1931 Landau and Peierls \cite{LandauPeierls1931} claimed that \emph{no} nonlocal variables, not even momentum, could be measured instantaneously in the relativistic range.

However, there arguments were unclear. They used the uncertainty relation $\Delta E \Delta t \gtrsim \hbar$, which is obtained by inserting the de Broglie relation $p = h/\lambda$ into the bandwidth theorem from the classical theory of waves, and considered the energy of interaction between the system and the measuring device. They thereby obtained the relation $(v - v')\Delta p \gtrsim \hbar/\Delta t$, where $v$ and $v'$ supposedly represent the velocities of the particle before and after the measurement, in the nonrelativistic range. They then substituted $(v-v')_{max} = c$, in accordance with relativistic causality, to obtain $\Delta p \Delta t \gtrsim \hbar/c$, which may also be obtained by substituting $E \approx pc$ for a relativistic particle into $\Delta E \Delta t \gtrsim \hbar$. Landau and Peierls argued that, as $(v - v')$ may be made arbitrarily large in nonrelativistic quantum theory, the value of $p$ could be determined precisely. They therefore claimed that, in the relativistic regime, ``the concept of momentum has a precise significance only over long times" and that similar arguments applied to other nonlocal properties. In effect they claimed that no instantaneous measurement of a nonlocal variable was possible in relativistic quantum mechanics. However, this idea seems to have rested upon a misinterpretation of their own equation. In the relation $\Delta E \Delta t \gtrsim \hbar$, the $\Delta t$ refers to the temporal extent of a wave packet describing the state of a system. In the relativistic regime, this is given by $\Delta t = \Delta x/c$, where $\Delta x$ is the spatial extent, or width. For example, if $\Delta t$ is the temporal extent of a photon pulse from a laser, then $\Delta E$ is the statistical spread in the observed energies of the individual photons detected. Equivalently, $\Delta t$ is the statistical spread in the observed
times at which the individual photons reach the detector. Similar arguments apply to other quantum mechanical particles. Under no circumstances is $\Delta t$ the ``time taken to perform a measurement of E", as claimed in \cite{LandauPeierls1931}, or is $\Delta E$ the ``uncertainty in a possessed value of E". In all measurements of energy, at all times, the system can be said to possess a particular value of $E$ at that particular time. This was in fact noted by Landau and Peierls in an earlier section of the same paper and the correct interpretation was attributed, by them, to Bohr \cite{Bohr1928}. 

It is therefore odd that they should have seemed to consider $\Delta t$ as the time taken to perform a measurement, and $\Delta p$ as the corresponding uncertainty in the \emph{possessed} value of $p$, when considering the relativistic range. It is also true that they considered the system as \emph{necessarily} possessing a momentum $p'$ before the momentum measurement, as well as a momentum $p$ after the measurement, when deriving the formula for the nonrelativistic case. This is also at odds with the standard interpretation of quantum mechanics, although the formula for the relativistic range, $\Delta p \Delta t \gtrsim \hbar/c$, remains valid.

In any event it was, until recently, by no means clear whether any instantaneous measurements of nonlocal variables, excluding functions of $p$, were possible, although it was generally, tacitly, assumed that they were not. That was until 1981 when Yakir Aharonov of Tel Aviv University and David Z. Albert of the Rockefeller University in New York proposed the first experiment designed specifically to measure a nonlocal property of a composite system \cite{AharonovAlbert1981}.

%%%%%%%%%%%%%%%%%%%%%%%%%%%%%%%%%%%%%%%%%%%%%%%%%%%%%%%
%%%%%%%%%%%%%%%%%%%%%%%%%%%%%%%%%%%%%%%%%%%%%%%%%%%%%%%
\section{Fifty years later, a breakthrough - ``Yes"}%``
\label{SectIV}

%%%%%%%%%%%%%%%%%%%%%%%%%%%%%%%%%%%%%%%%%%%%%%%%%%%%%%%
\subsection{``Yes - at least in some cases" - Aharonov and Albert (1981)}%``
\label{SectIV.1}

For simplicity Aharonov and Albert considered the two part system discussed in the previous section as an example of how to perform a measurement of a nonlocal variable without violating causality. They designed an experiment to measure $\hat{\sigma}_z$, the $z$-component of the total spin of one of the maximally entangled Bell states, $\Ket{\Psi}_{+}$ or $\Ket{\Psi}_{-}$, without measuring either $\hat{\sigma}_z^{(A)}$ or $\hat{\sigma}_z^{(B)}$ individually. They referred to this as a nondemolition experiment, as it left the nonlocal state intact and did not trigger the reduction of the state vector, which would involve the loss on the nonlocal property. The general idea was to get the nonlocal variable of the entangled system to couple to local variables at different parts of the measuring apparatus. The results of these local measurements could then be combined via classical information transfer to reveal the value of the nonlocal property without violating causality or causing state vector reduction. Aharonov and Albert first considered a piece of apparatus designed to measure the $z$-component of the spin of a single particle, say $\hat{\sigma}_z^{(A)}$. The device interacts with the one-particle system over a short period via the interaction Hamiltonian,
\begin{eqnarray} \label{H_intA}
\hat{H}_{int}^{(A)} = g_A(t)q_A(t)\hat{\sigma}_z^{(A)},
\end{eqnarray}
where $g_A(t)$ represents the coupling between the device and the system, and is nonzero only during the interval $t_0 < t < t_0 + \epsilon$, and $q_A(t)$ is an internal variable associated with the measuring device. The particle then gains momentum $\Pi_A$, which is the momentum canonically conjugate to $q_A$, given via
\begin{eqnarray} \label{Pi_A}
\frac{\partial \Pi_A}{\partial t} = -g_A(t)\hat{\sigma}_z^{(A)}.
\end{eqnarray}
The simplest and most common example of such a device is the Stern-Gerlach apparatus, which can be used to measure the $z$-component of spin for charged particles such as electrons. In this case, the coupling $g_A (t)$ is produced by the magnetic field, the internal variable $q_A$ is the $z$-component of the electron position and the momentum $\Pi_A$, conjugate to $q_A$, is the momentum that the electron acquires in the $z$-direction.

For a two particle system, using two such devices to perform simultaneous measurements at $x_A$ and $x_B$ would measure $\hat{\sigma}_z^{(A)}$ and $\hat{\sigma}_z^{(B)}$, respectively, necessarily causing the collapse of an entangled state. However, it is possible to modify the design of the measurement process in the individual devices in such a way that the interaction Hamiltonian retains the same form, but which allows the initial states of the two devices to be correlated. 
%\footnote{Many thanks to Dr. Berry Groisman for proposing this specific example.}
Such devices may then be used to perform a measurement of a nonlocal variable on the entangled two-particle state. We may use our Stern-Gerlach devices, each together with an additional ancillary electron, initially with spin up, to measure the individual spins of two particles in a slightly more complicated way. The coupling between the device and the particle is now the controlled not (or CNOT), coupling between the electron whose spin we wish to determine, which acts as the control, and the ancillary electron, which acts as the target. If the control electron initially had spin up, then the ancillary electron remains in the up state, and if the control electron originally had spin down, the spin of the ancillary electron will flip to down. The spin measurement in the Sten-Gerlach apparatus is then performed on the ancillary electron, from which we are able to infer the spin of the original control.

Again, this more complicated experiment may be used to determine the values of $\hat{\sigma}_z^{(A)}$ and $\hat{\sigma}_z^{(B)}$ individually. However, if the two modified Stern-Gerlach devices are first brought together so that the two ancillary electrons are allowed to interact and become entangled, it is possible to prepare the devices in an initial state so that,
\begin{subequations}
\begin{align}
\Pi_A(t_0) = -\Pi_B(t_0) \implies \Pi_A(t_0) + \Pi_B(t_0) = 0, \label{Pi.a}\\
q_A(t_0) = q_B(t_0) = q(t_0) \implies q_A(t_0) - q_B(t_0) = 0. \label{Pi.b}
\end{align}
\end{subequations}
Rearranging equations (\ref{H_intA}) and (\ref{Pi_A}) we see that,
\begin{eqnarray} \label{spinA}
\hat{\sigma}_z^{(A)} = \frac{\Pi_A(t_0)-\Pi_A(t_0+\epsilon)}{\int_{t=0}^{t_0+\epsilon}g(t)dt}, 
\end{eqnarray}
and that an analogous expression holds for $\hat{\sigma}_z^{(B)}$. The two devices are again placed at $x_A$ and $x_B$ but are now effectively one measuring device which acts on the two particle system via the interaction Hamiltonian
\begin{eqnarray} \label{}
\hat{H}_{int} = \hat{H}_{int}^{(A)} + \hat{H}_{int}^{(B)} = g(t)\left(q_A(t)\hat{\sigma}_z^{(A)} + q_B(t)\hat{\sigma}_z^{(B)}\right),
\end{eqnarray}
if $g(t) = g_A(t) = g_B(t)$. In addition, $q_A(t)$ will be equal to $q_B(t)$ at all times if their initial values are equal, according to Eq. (\ref{Pi.b}), and the couplings $g_A(t)$ and $g_B(t)$ are equal, so that the interaction Hamiltonian is given by
\begin{eqnarray} \label{}
\hat{H}_{int} = -q(t)q(t)\left(\frac{\Pi_A(t_0)+\Pi_B(t_0+\epsilon)}{\int_{t=0}^{t_0+\epsilon}g(t)dt}\right),
\end{eqnarray}
where $t_0 < t < t_0 + \epsilon$, in accordance with Eqs. (\ref{Pi.b}) and (\ref{spinA}). The apparatus has therefore measured the value of the nonlocal variable $\hat{\sigma}_z$, where
\begin{eqnarray} \label{}
\hat{\sigma}_z = \hat{\sigma}_z^{(A)} + \hat{\sigma}_z^{(B)} = -\left(\frac{\Pi_A(t_0+\epsilon)+\Pi_B(t_0+\epsilon)}{\int_{t=0}^{t_0+\epsilon}g(t)dt}\right),
\end{eqnarray}
without determining either $\hat{\sigma}_z^{(A)}$ or $\hat{\sigma}_z^{(B)}$ individually, and so without destroying the nonlocal state. Furthermore, relativistic causality is not violated as the conjugate momentums of the two ancillary electrons, given by $\Pi_A(t_0+\epsilon)$ and  $\Pi_B(t_0+\epsilon)$ are measured using only local interactions. Although the result given by the macroscopic part of one of the measuring devices, say that at $x_A$, such as the position of the pointer or dials that records the conjugate momentum and spin of the ancillary electron, depend upon which state the ancillary wave function has collapsed to, it is not \emph{causally} dependent upon the result obtained at $x_B$ because the dial at $x_B$ will read $\pm \hbar/2$, randomly.

%%%%%%%%%%%%%%%%%%%%%%%%%%%%%%%%%%%%%%%%%%%%%%%%%
\subsection{Lorentz invariance}%``
\label{SectIV.2}

It is also worth noting that $\epsilon$ may be made arbitrarily small, so that each local interaction is effectively instantaneous, and that the instantaneous measurements of $\Pi_A$ and $\Pi_B$ need not be simultaneous. It can be shown that if $\Pi_A$ is recorded at time $t_1$ and $\Pi_B$ at $t_2>t_1$, then, during the interval between the two measurements, the system will not be in an eigenstate of any operator. In fact, the full state will not be any direct product of a state of the two-particle system and a state of the apparatus. The second interaction at $x_B$ then ``undoes" the disturbance caused by the first interaction and restores the initial nonlocal state of the system together with some \emph{new} local state of the apparatus in which the value of a nonlocal property is encoded. Although a finite amount of time is required to combine the results of the local measurements, the measurement itself is instantaneous as the value of $\hat{\sigma}_z$ becomes encoded in the local properties of the apparatus immediately after the final interaction. This is true in any inertial frame, regardless of which interaction the observer sees \emph{first}. 

The experimental probabilities are therefore, remarkably, completely Lorentz invariant, despite the non invariance of the state history under Lorentz boosts. This led Aharonov and Albert to the conjecture that ``the covariance of relativistic quantum theories resides \emph{exclusively} in the experimental probabilities and not in the underlying quantum states" \cite{AharonovAlbert1981}.

%%%%%%%%%%%%%%%%%%%%%%%%%%%%%%%%%%%%%%%%%%%%%%%%%%%``
\subsection{Additional considerations}%``
\label{SectIV.3}

Aharonov and Albert also produced a number of other interesting results regarding causal measurements of non-local variables in their 1981 paper. In addition to the experiment outlined above they also proposed what they called a \emph{state specific verification measurement} to verify the value of $\hat{\sigma}^{2}$ \emph{if} $\hat{\sigma}^{2} = 0$, or, in other words, to verify that the state of the system is $\Ket{\Psi} = \Ket{\Psi_{-}}$. This type of measurement is more limited than a measurement of $\hat{\sigma}^{2}$ in the usual sense, which Aharonov and Albert called an \emph{operator specific} measurement, in that, if $\hat{\sigma}^{2} \neq 0$, the interaction destroys the initial state and we are unable to determine what value of $\hat{\sigma}^{2}$ it originally possessed.
It is possible to design a state verification measurement for $\Ket{\Psi_{-}}$ by noting that the state is completely specified by $\Ket{\hat{\sigma}^{2}=0}$ in our bipartite system. Equivalently,
\begin{subequations}
\begin{align}
&\hat{\sigma}_x = \hat{\sigma}_x^{(A)} + \hat{\sigma}_x^{(B)} = 0, \label{const.a} \\
&\hat{\sigma}_y = \hat{\sigma}_y^{(A)} + \hat{\sigma}_y^{(B)} = 0, \label{const.b} \\
&\hat{\sigma}_z = \hat{\sigma}_z^{(A)} + \hat{\sigma}_z^{(B)} = 0. \label{const.c}
\end{align}
\end{subequations}
We may then construct a measuring device which will interact with the system via the Hamiltonian 
\begin{eqnarray} \label{}
\hat{H}_{int} = g(t)\left(\hat{\sigma}_x^{(A)}q_x^{(A)} + \hat{\sigma}_y^{(A)}q_y^{(A)} + \hat{\sigma}_z^{(A)}q_z^{(A)} + \hat{\sigma}_x^{(B)}q_x^{(B)} + \hat{\sigma}_y^{(B)}q_y^{(B)} + \hat{\sigma}_z^{(B)}q_z^{(B)}\right).
\end{eqnarray}
This may be done using three sets of Stern-Gerlach equipment and three pairs of ancillary electrons in a manner analogous to that outlined previously. The devices may then be prepared so that,
\begin{subequations}
\begin{align}
&q_x^{(A)} = q_x^{(B)}, \ \ q_y^{(A)} = \ \ q_y^{(B)}, \ \ q_z^{(A)} = q_z^{(B)}, \label{qAqB} \\
&\Pi_x^{(A)} = -\Pi_x^{(B)}, \ \ \Pi_y^{(A)} = -\Pi_y^{(B)}, \ \ \Pi_z^{(A)} = -\Pi_z^{(B)}, \label{PiAPiB}
\end{align}
\end{subequations}
for each of the three Stern-Gerlach ancillary electron pairs. Using equations (\ref{qAqB})-(\ref{PiAPiB}), the Hamiltonian may be written in the form 
\begin{eqnarray} \label{}
\hat{H}_{int} = g(t)\left[\left(\hat{\sigma}_x^{(A)} + \hat{\sigma}_x^{(B)}\right)\left(q_x^{(A)} + q_x^{(B)}\right) + \left(\hat{\sigma}_y^{(A)} + \hat{\sigma}_y^{(B)}\right)\left(q_y^{(A)} + q_y^{(B)}\right) + \left(\hat{\sigma}_z^{(A)} + \hat{\sigma}_z^{(B)}\right)\left(q_z^{(A)} + q_z^{(B)}\right)\right]^{\frac{1}{2}},
\end{eqnarray}
from which the equations of motion for $\left(\hat{\sigma}_x^{(A)} + \hat{\sigma}_x^{(B)}\right)$,  $\left(\hat{\sigma}_y^{(A)} + \hat{\sigma}_y^{(B)}\right)$ and $\left(\hat{\sigma}_z^{(A)} + \hat{\sigma}_z^{(B)}\right)$ may be shown to be,
\begin{subequations}
\begin{align}
&\partial_{t} \left(\hat{\sigma}_x^{(A)} + \hat{\sigma}_x^{(B)}\right) = g(t)\left[\left(q_y^{(A)} + q_y^{(B)}\right)\left(\hat{\sigma}_z^{(A)} + \hat{\sigma}_z^{(B)}\right)-\left(q_z^{(A)} + q_z^{(B)}\right)\left(\hat{\sigma}_y^{(A)} + \hat{\sigma}_y^{(B)}\right)\right]^{\frac{1}{2}}, \label{DiffEq.a} \\ 
%&\partial_{t} \left(\hat{\sigma}_y^{(A)} + \hat{\sigma}_y^{(B)}\right) = g(t)\left[\left(q_z^{(A)} + q_z^{(B)}\right)\left(\hat{\sigma}_y^{(A)} + \hat{\sigma}_y^{(B)}\right)-\left(q_x^{(A)} + q_x^{(B)}\right)\left(\hat{\sigma}_z^{(A)} + \hat{\sigma}_z^{(B)}\right)\right]^{\frac{1}{2}}, \label{DiffEq.b} \\ 
&\partial_{t} \left(\hat{\sigma}_y^{(A)} + \hat{\sigma}_y^{(B)}\right) = g(t)\left[\left(q_z^{(A)} + q_z^{(B)}\right)\left(\hat{\sigma}_x^{(A)} + \hat{\sigma}_x^{(B)}\right)-\left(q_x^{(A)} + q_x^{(B)}\right)\left(\hat{\sigma}_z^{(A)} + \hat{\sigma}_z^{(B)}\right)\right]^{\frac{1}{2}}, \label{DiffEq.b} \\ 
&\partial_{t} \left(\hat{\sigma}_z^{(A)} + \hat{\sigma}_z^{(B)}\right) = g(t)\left[\left(q_x^{(A)} + q_x^{(B)}\right)\left(\hat{\sigma}_y^{(A)} + \hat{\sigma}_y^{(B)}\right)-\left(q_y^{(A)} + q_y^{(B)}\right)\left(\hat{\sigma}_x^{(A)} + \hat{\sigma}_x^{(B)}\right)\right]^{\frac{1}{2}}. \label{DiffEq.c} 
\end{align}
\end{subequations}
It is then straightforward to see that the constant functions (\ref{const.a})-(\ref{const.c}) are solutions of the differential equations (\ref{DiffEq.a})-(\ref{DiffEq.c}), and that such an experiment would enable us to verify the existence of the state $\Ket{\Psi_{-}} = \Ket{\hat{\sigma}_{z}=0, \hat{\sigma}^2=0}$ by means of local interactions, without destroying it, and without violating causality.

Therefore, performing the \emph{operator specific nondemolition experiment} to measure the value of $\hat{\sigma}_{z}$, we obtain $\hat{\sigma}_{z}=0$ if our state is one of the maximally entangled Bell states. If this is followed by a \emph{state specific verification experiment} for $\Ket{\Psi_{-}}$, we may determine the value of $\hat{\sigma}^2$ to be either $0$ or $+2\hbar^2$. We have, therefore, effectively measured the values of two nonlocal variables although, in the later case when $\hat{\sigma}^2 = +2\hbar^2$, the initial state is destroyed.

In any event, Aharonov and Albert's thought experiment showed conclusively that at least some nonlocal properties of physical systems can be measured in relativistic quantum theory. This was a tremendous breakthrough which took half a century to be realised. However, it naturally raised many important questions. For our example system the obvious question is ``Is it possible to verify \emph{every} linear combination of the local states $\Ket{\Psi_{1}}$-$\Ket{\Psi_{4}}$?". This question was also considered in their 1981 paper and it was found that an arbitrary state of the form
\begin{eqnarray} \label{}
\Ket{\Psi} = \alpha\Ket{\Psi_{a}} \otimes \Ket{\Psi_{b}} + \beta\Ket{\Psi_{c}} \otimes \Ket{\Psi_{d}},
\end{eqnarray}
where $a,b,c,d \in \left\{1,2,3,4\right\}$ and $\alpha,\beta$ are arbitrary complex coefficients, could \emph{not} be verified instantaneously by means of a nondemolition experiment like the ones proposed without violating causality. This was established by means of a specific counter example, considering the state 
\begin{eqnarray} \label{}
\Ket{\Psi_{\phi}} = \sin\phi \Ket{\Psi_{4}} + \cos\phi \Ket{\Psi_{3}} = \sin\phi \Ket{\downarrow_z}_A\Ket{\uparrow_z}_B + \cos\phi \Ket{\uparrow_z}_A\Ket{\downarrow_z}_B.
\end{eqnarray}
It was shown that any instantaneous nondemolition experiment distinguishing between $\Ket{\Psi_{\phi}}$ and a perpendicular state $\Ket{\Psi_{\phi_{\perp}}} =  \sin\phi \Ket{\Psi_{\phi}} - \sin\phi \Ket{\Psi_{4}}$ must necessarily violate causality unless $\phi$ took the particular values $\phi = n\pi/4$, $n \in \mathbb{N}$. Aharonov and Albert showed that any experiment capable of distinguishing between $\Ket{\Psi_{\phi}}$ and $\Ket{\Psi_{\phi_{\perp}}}$ necessarily involves the measurement of some observable $M_{\phi}$ of which $\Ket{\Psi_{\phi}}$ is an eigenstate $\left(M_{\phi}\Ket{\Psi_{\phi}} = \gamma \Ket{\Psi_{\phi}}\right)$, and for which any eigenstate degenerate with $\Ket{\Psi_{\phi}}$ is orthogonal to $\Ket{\Psi_{\phi_{\perp}}}$. They then proposed the following two scenarios.\\
\\
\emph{Scenario (1)}

\begin{itemize} 

\item $t<< t_0$, the system is initially prepared in state $\Ket{\Psi_{-}}$, ($\phi = \pi/4$)

\item $t = t_0$, a measurement of the variable $M_{\phi}$ is made

\item $t = t_0 + \epsilon$, a local measurement of one of the fermion spins, say $\hat{\sigma}_x^{(B)}$, is performed

\end{itemize}

\noindent
\emph{Scenario (2)}

\begin{itemize} 

\item $t<< t_0$, the system is initially prepared in state $\Ket{\Psi_{-}}$, ($\phi = \pi/4$)

\item $t = t_0 - \epsilon$, the state of the system is changed from $\Ket{\Psi_{-}}$ to $\Ket{\Psi_{+}}$, for example by rotating the $z$-spin of the particle at $x_A$ using a magnetic field

\item $t = t_0$, a measurement of the variable $M_{\phi}$ is made

\item $t = t_0 + \epsilon$, a local measurement of $\hat{\sigma}_x^{(B)}$ is made.

\end{itemize}

It is straightforward, but time consuming, to show that the probabilities of obtaining $\hat{\sigma}_x^{(B)}=+\hbar/2$ for scenarios (1) and (2) are given by 
\begin{subequations}
\begin{align}
&P^{(1)}\left(\hat{\sigma}_z^{(B)}=+\hbar/2\right) = \left[(1-2\sin\phi \cos\phi)\cos^2\phi+(1+2\sin\phi \cos\phi)\sin^2\phi + \eta\right]^{\frac{1}{2}}, \label{} \\ 
&P^{(2)}\left(\hat{\sigma}_z^{(B)}=+\hbar/2\right) =  \left[(1+2\sin\phi \cos\phi)\cos^2\phi+(1-2\sin\phi \cos\phi)\sin^2\phi + \eta\right]^{\frac{1}{2}}, \label{}
\end{align}
\end{subequations}
respectively, where $\eta$ is a function of the Eulerian angles of $M_{\phi}$. The proof shall therefore be omitted here and the interested reader is referred to \cite{AharonovAlbert1981}. These two equations are, in general, not equal and so the probabilities of the outcomes of local interactions at $x_B$ will depend \emph{causally} upon conditions at $x_A$ and vice-versa unless $\eta$ satisfies certain constraints. For probabilities at $x_B$ to be independent of conditions at $x_A$ we require $\eta = \cos^2\phi - \sin^2\phi$, and for probabilities at $x_A$ to be independent of conditions at $x_B$ we require $\eta = \sin^2\phi - \cos^2\phi$. Causality therefore requires that,
\begin{eqnarray} \label{}
\eta = \cos^2\phi - \sin^2\phi = \sin^2\phi - \cos^2\phi = 0 %\implies \phi = \frac{n\pi}{4}, \ n \in \mathbb{N}.
\end{eqnarray}
or, in other words, $\phi = n\pi/4$, $n \in \mathbb{N}$.

%%%%%%%%%%%%%%%%%%%%%%%%%%%%%%%%%%%%%%%%%%%%%%%%%%%``
\subsection{Unanswered questions}
\label{SectIV.4}

Thus the work of Aharonov and Albert in 1981, although groundbreaking, left open two important questions,

\begin{enumerate}

\item  \emph{What is the largest class of nonlocal states whose nonlocal properties may be verified by means of nondemolition experiments like the ones outlined above?}

\item \emph{Are there any other forms of measurement, nondemolition or otherwise, by which a larger class of nonlocal variables can be measured?}

\end{enumerate}

It was therefore necessary in subsequent work to consider carefully what was meant by the term ``measurement". The rest of this essay is devoted mainly to considering the results of four important papers which, following on from Aharonov and Albert's original discoveries, attempt to answer the questions above \cite{AharonovAlbertVaidman1986,GroismanReznik2002,PopescuVaidman1994,AharonovVaidman2000}. The first of these, written by Aharonov and Albert together with Lev Vaidman, also of Tel Aviv University, was published in 1986 \cite{AharonovAlbertVaidman1986}. This generalised their earlier results to composite systems of an arbitrary number of subsystems and succeeded in establishing exactly what classes of nonlocal operators could be measured by means
of nondemolition experiments involving only local interactions. These results were
then used to show what classes of nonlocal states were verifiable by such procedures.

%%%%%%%%%%%%%%%%%%%%%%%%%%%%%%%%%%%%%%%%%%%%%%%%%%%``
%%%%%%%%%%%%%%%%%%%%%%%%%%%%%%%%%%%%%%%%%%%%%%%%%%%``
\section{But in which cases exactly? - ``Quite a lot, but not all, at least, not using our original procedure" - Aharonov, Albert and Vaidman (1986)}
\label{SectV}

The 1986 paper by Aharonov, Albert and Vaidman \cite{AharonovAlbertVaidman1986} considered a general composite quantum system composed of an arbitrary number, $N$, of spatially separated subsystems, as shown in Fig. 1.
\begin{figure} \label{MR} 
%\sidecaption
 \begin{center}
\psfig{file=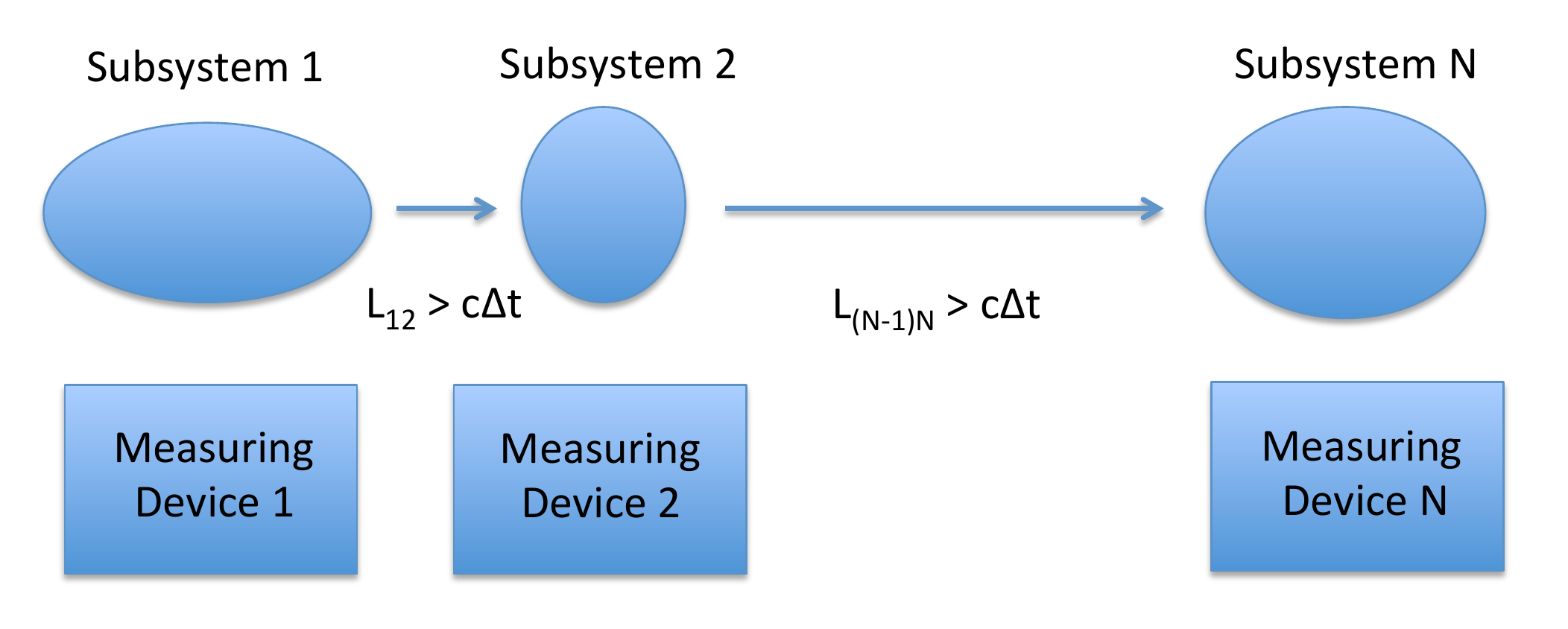,width=5.5in}
\caption{Space-like separated subsystems with local measuring devices}      
 \end{center}
\end{figure}
It was assumed that any operator $\hat{A}_i$, representing a local property of one of the subsystems, could be measured, and an experiment to measure any nonlocal operator of the form 
\begin{eqnarray} \label{SumOp}
\sum_{i=1}^{N}\hat{A}_i
\end{eqnarray}
was devised. For a general composite system the measuring device consists of $N$ spatially separate parts, each of which interacts with the system via the Hamiltonian,
\begin{eqnarray} \label{}
\hat{H}_{int}^{(i)} = g(t)q_i(t)\hat{A}_i.
\end{eqnarray}
The composite device is then interacts via the Hamiltonian 
\begin{eqnarray} \label{H_int*}
\hat{H}_{int} = g(t)\sum_{i=1}^{N}q_i(t)\hat{A}_i,
\end{eqnarray}
and is prepared in the initial state given by 
\begin{subequations}
\begin{align}
&(q_i - q_j)_{t=t_0} = 0, \ \forall i,j = 1,2 . . N, \label{} \\
&\sum_{i=1}^{N}\Pi_i\bigg|_{t=t_0} = 0, \label{SumZero}
\end{align}
\end{subequations}
by analogy with the simpler $N = 2$ case. The device then interacts with the whole system over the interval $t_0 < t < t_0 + \epsilon$ such that the normalisation condition,
\begin{eqnarray} \label{Norm}
\int_{t=t_0}^{t_0+\epsilon}g(t)dt = 1,
\end{eqnarray}
is fulfilled. During the interaction, the momenta canonically conjugate to the local degrees of freedom $\left\{q_i\right\}$ obey the equation 
\begin{eqnarray} \label{}
\dot{\Pi}_i = -g(t)\hat{A}_i,
\end{eqnarray}
in the Heisenberg picture. From this, it is simple to demonstrate that the sum of the results of local measurements of $\Pi_i$ is equal to the value of the nonlocal variable we wish to measure. Thus, we have
\begin{eqnarray} \label{}
%\sum_{i=1}^{N}\dot{\Pi}_i = -g(t)\sum_{i=1}^{N}\hat{A}_i \implies 
\left[\sum_{i=1}^{N}\Pi_i \right]_{t_0}^{t_0+\epsilon} = -\int_{t_0}^{t_0+\epsilon}g(t)dt \sum_{i=1}^{N}\hat{A}_i,
\end{eqnarray}
so that 
\begin{eqnarray} \label{}
\sum_{i=1}^{N}\hat{A}_i = -\sum_{i=1}^{N}\Pi_i \bigg|_{t=t_0+\epsilon},
\end{eqnarray}
where the last step follows from Eqs. (\ref{Norm}) and (\ref{SumZero}). Again we have initially considered simultaneous measurements over a finite period, but the time interval may be made arbitrarily small and the instantaneous measurements may be performed by space-like separated observers at different times, without altering the probabilities of local measurements.

Thus we may instantaneously measure \emph{any} nonlocal variables of the form Eq. (\ref{SumOp}) possessed by any nonlocal state of a composite quantum system with $N$ subsystems, in a nondemolition experiment involving only local interactions. The local probabilities of our experiment will remain unchanged, even if our local measurements are not simultaneous, as will certainly be the case in all but \emph{one} inertial frame. The measurement does not violate relativistic causality even if the space-like separation between different subsystems $L_{i j}$ is such that $L_{i j} = |x_j-x_i| \gg c\Delta t$, where $\Delta t$ is the time between measurements at $x_i$ and $x_j$ in \emph{any} inertial frame.

By setting $\hat{A}'_i = \alpha_i \hat{A}_i$ and measuring $\sum_{i=1}^{N}\hat{A}'_i$ as before, we may also measure any linear
sum of local operators 
\begin{eqnarray} \label{class1}
\sum_{i=1}^{N}\alpha_i\hat{A}_i.
\end{eqnarray}
Alternatively, by setting $\hat{A}''_i = \ln(\hat{A}_i)$ and taking the exponent of $\sum_{i=1}^{N}\hat{A}''_i$, we may measure any product of local operators via 
\begin{eqnarray} \label{class2}
\prod_{i=1}^{N}\hat{A}_i = \exp\left(\sum_{i=1}^{N}\ln(\hat{A}_i)\right).
\end{eqnarray}
A third class on nonlocal variables, the modular sums of $i$ local operators,
\begin{eqnarray} \label{class3}
\left[\sum_{i=1}^{N}\hat{A}_i\right]mod(a),
\end{eqnarray}
where $a$ is an arbitrary constant, may also be measured if the initial state of the measuring apparatus is modified such that,
\begin{subequations}
\begin{align}
&(q_i - q_j)_{t=t_0} = 0, \ \forall i,j = 1,2 . . N, \label{} \\
&\left[\sum_{i=1}^{N}\Pi_i\right]mod(a)\bigg|_{t=t_0} = 0, \ q_i mod\left(\frac{2\pi\hbar}{a}\right)\bigg|_{t=t_0} = 0. \label{SumZeroMod}
\end{align}
\end{subequations}
Using exactly the same procedure as before this now gives,
\begin{eqnarray} \label{}
\left[\sum_{i=1}^{N}\hat{A}_i\right]mod(a) = -\left[\sum_{i=1}^{N}\Pi_i\right]mod(a)\bigg|_{t=t_0+\epsilon}.
\end{eqnarray}
It can then be shown that all such sums must be equal to zero for all values of $a$ i.e. that our measurement procedure is equivalent to a nondemolition verification that,
\begin{eqnarray} \label{SumZeroOp*}
\left[\sum_{i=1}^{N}\hat{A}_i\right]mod(a) = 0.
\end{eqnarray}
This can be seen by noting that condition 
\begin{eqnarray} \label{}
\exp\left(-\frac{i}{\hbar}\int_{t_0}^{t_0+\epsilon}\hat{H}_{int}dt\right)\Ket{\Psi_{in}} = \Ket{\Psi_{in}},
\end{eqnarray}
which states that the time translation operator, acting on the measuring apparatus and the initial state of the system
$\Ket{\Psi_{in}}$ during the measurement, does not change $\Ket{\Psi_{in}}$, is equivalent to condition
\begin{eqnarray} \label{}
\left[-\frac{1}{\hbar}\int_{t_0}^{t_0+\epsilon}\hat{H}_{int}dt\right]mod(2\pi)\Ket{\Psi_{in}} = 0.
\end{eqnarray}
Substituting in for $\hat{H}_{int}$ int from (\ref{H_int*}) and using (\ref{Norm}) together with the fact that all, during the interaction, $q_i = q_j = q$, $\forall i,j$, we get
\begin{eqnarray} \label{}
\left[-\frac{1}{\hbar}q\sum \hat{A}_i\right]mod(2\pi)\Ket{\Psi_{in}} = 0.
\end{eqnarray}
Using (\ref{SumZeroMod}) we can then see that this equation is satisfied by (\ref{SumZeroOp*}) and that the modular sum is equal to zero.

In their later work therefore, Aharonov and Albert found three general classes of nonlocal variables which could be measured instantaneously using appropriate extensions of their original procedure. However, there was no reason to assume that it was
possible to measure \emph{any} arbitrary function of local variables using this method. Although question (2) from Sect. \ref{SectIV.4} remained open, question (1) could now be answered with reasonable certainty. A nondemolition verification of a nonlocal state $\Ket{\Psi}$ involves \emph{enough} nondemolition measurements of nonlocal variables to \emph{specify} $\Ket{\Psi}$. Using only the types of nondemolition experiments considered so far, it is possible to show that, for composite systems of $M$, $K$-dimensional, subsystems, only states of the form
\begin{eqnarray} \label{states}
\Ket{\Psi} = \frac{1}{\sqrt{K}}\sum_{i=1}^{K}\Ket{i}_1 \otimes \Ket{i}_2 \otimes . . . \Ket{i}_M,
\end{eqnarray}
may be measured. The proof of this statement is in three parts. Firstly, it is necessary to show that \emph{any} nonlocal state of such a system may be written in \emph{canonical form},
\begin{eqnarray} \label{PhiCanon}
\Ket{\Psi} = \sum_{i=1}^{K}\alpha_{i}\Ket{i}_1 \otimes \Ket{i}_2 \otimes . . . \Ket{i}_M,
\end{eqnarray}
where $\alpha_{i} \neq 0$, by an appropriate choice of basis vectors in each of the subsystems. Secondly, we must show that states of the form (\ref{PhiCanon}) are measurable using the procedures above. In the third and final step we show that if our system is divided in any way in to two nonempty subsystems, and the basis of the state spaces of these two subsystems are chosen such that the state of the composite system is in canonical form (\ref{PhiCanon}), then the coefficients $\left\{\alpha_i\right\}$ must necessarily all be equal. This step, in conjunction with the first two, therefore demonstrates that \emph{only} nonlocal states that have canonical form, in which all the coefficients are equal, may be verified by the methods considered so far. We now outline the essential details of the proof given in \cite{AharonovAlbertVaidman1986}.\\
\\ 
\indent
\emph{Stage (1)}: The proof is quite straightforward but, for brevity. we will assume that any nonlocal state may be written in canonical form, and concern ourselves with the last two propositions.

\emph{Stage (2)}: We may verify a nonlocal state of the form (\ref{PhiCanon}) by a measurement procedure in two stages. The first stage consists of $M-1$ measurements which verify that the sum of two local operators, $\hat{A}_1$ and $\hat{A}_l$, is $\hat{A}_1 + \hat{A}_l = 0$, where $\hat{A}_1$ and $\hat{A}_l$ are given by
\begin{subequations}
\begin{align}
&\hat{A}_1\Ket{i}_1 = -i\Ket{i}_1 \label{} \\
&\hat{A}_l\Ket{i}_l = +i\Ket{i}_l, \ l=2,3 . . M. \label{} 
\end{align}
\end{subequations}
This is equivalent to verifying that the state $\Ket{\Psi}$ has canonical form in \emph{some} basis, without specifying the values of the coefficients $\left\{\alpha_i\right\}$. In the second stage we start by defining a set of unitary operators $\left\{\hat{U}_m\right\}$, $m=1,2 . . M$, which act in every local subsystem such that,
\begin{subequations}
\begin{align}
&\hat{U}_m\Ket{i}_m = \Ket{i+1}_m, \ i=1,2, . . .(K-1) \label{} \\
&\hat{U}_m\Ket{K}_m = \Ket{K}_m, \ m=1,2 . . M,\label{} 
\end{align}
\end{subequations}
and which therefore satisfy
\begin{eqnarray} \label{ProdInvar}
\prod_{m=1}^{M}\hat{U}_m\Ket{\Psi} = \Ket{\Psi}.
\end{eqnarray}
Then, by defining another set of local operators $\left\{B_m\right\}$, $m=1,2 . . M$, where
\begin{eqnarray} \label{}
\hat{U}_m = e^{i\hat{B}_m},
\end{eqnarray}
we see that (\ref{ProdInvar}) is equivalent to 
\begin{eqnarray} \label{}
\left[\sum_{m=1}^{M}\hat{B}_m\right]mod(2\pi) = 0,
\end{eqnarray}
which we are able to verify as shown above. This completes the second stage of the proof. The third part of the proof is more
complicated and, for simplicity, we will consider the case of a bipartite system in detail, before stating the general result.\\
\\
\indent
\emph{Stage (3)}: It can be shown that, in order to be verified without violating causality, our state $\Ket{\Psi}$ must be of the form
\begin{eqnarray} \label{Phi}
\Ket{\Psi} = \sum_{i,j}\beta_{ij}\Ket{i}_1 \otimes \Ket{j}_2,
\end{eqnarray}
where $\Ket{i}_1 \otimes \Ket{j}_2$ is one of the degenerate eigenstates of the nonlocal operator we wish to measure. If this is not the case it is possible send information at superluminal speeds between the two parts of the system. This may be illustrated by the following procedure. First, suppose we have three sets of eigenstates of a nonlocal operator $\Ket{i_1}_1 \otimes \Ket{j_1}_2$, $\Ket{i_1}_1 \otimes \Ket{j_2}_2$ and $\Ket{i_2}_1 \otimes \Ket{j_1}_2$, which are degenerate with each
other, and assume that there is a fourth, $\Ket{i_2}_1 \otimes \Ket{j_2}_2$, which is \emph{not} degenerate with the rest. Then the procedure is 

\begin{itemize}

\item $t << t_0$, prepare subsystem two in state $\Ket{\Psi}_2 = \alpha_1\Ket{j_1}_2 + \alpha_2\Ket{j_2}_2$, $\alpha_1,\alpha_2 \neq 0$

\item $t = t_0 - \epsilon$, prepare subsystem one in state $\Ket{\Psi}_1 = \Ket{i_1}_1$ or $\Ket{\Psi}_1 = \Ket{i_2}_1$

\item $t = t_0$, perform a measurement of the nonlocal variable

\item $t = t_0 + \epsilon$, perform a local verification measurement of state $\Ket{\Psi}_2$

\end{itemize}

If at $t = t_0 - \epsilon$, we prepare $\Ket{\Psi}_1 = \Ket{i_1}_1$ then the measurement of the nonlocal variable at $t = t_0$ will not change the state $\Ket{\Psi}_2$ and the result of the state verification at $t = t_0 - \epsilon$ will be ``Yes" with probability $1$. This is because $\Ket{\Psi} = \Ket{\Psi}_1 \otimes \Ket{\Psi}_2 = \alpha_1\Ket{i_1}_1 \otimes \Ket{j_1}_2 + \alpha_2\Ket{i_1}_1 \otimes \Ket{j_2}_2$ is also an eigenstate of the nonlocal operator and so $\Ket{\Psi}_2$ will not be disturbed.

If, on the other hand, at $t = t_0 - \epsilon$ we prepare $\Ket{\Psi}_1 = \Ket{i_2}_1$ the composite state will be given by $\Ket{\Psi} = \Ket{\Psi}_1 \otimes \Ket{\Psi}_2 = \alpha_1\Ket{i_2}_1 \otimes \Ket{j_1}_2 + \alpha_2\Ket{i_2}_1 \otimes \Ket{j_2}_2$ which is not an eigenstate of the nonlocal operator. Thus, the measurement at $t = t_0 + \epsilon$ will disturb the state and reveal either $\Ket{j_1}_2$ or $\Ket{j_2}_2$ with probabilities $|\alpha_1|^2$ and $|\alpha_2|^2$, respectively. It is therefore possible for local interactions in subsystem one to affect the probabilities of results of local measurements of subsystem two at time $2 \epsilon$ later. As $ \epsilon \rightarrow 0$, causality is violated.

The matrix elements of equation (\ref{Phi}) are therefore nonzero if and only if the state $\Ket{i}_1 \otimes \Ket{j}_2$ is degenerate and the measurement of our nonlocal operator is equivalent to a verification that the matrix $\beta$ may be written in block diagonal form,
\[\beta = \left[\begin{array}{cc} \label{}
\beta'   &    0   \\
  0        &     0
\end{array} \right], \] 
where $\dim(\beta')$ is equal to the number of degenerate eigenstates, by reordering the local basis vectors appropriately. We also know that the state $\Ket{\Psi}$ may be written in canonical form (\ref{PhiCanon}) where the matrix $\alpha$ is nonsingular. By equating these two results we obtain,
\begin{eqnarray} \label{}
\Ket{\Psi} = \sum_{i,j}\beta_{ij}\Ket{i}_1 \otimes \Ket{j}_2 = \sum_{i=1}^{K}\alpha_i \Ket{i}_1 \otimes \Ket{i}_2,
\end{eqnarray}
which is equivalent to stating that $\beta$ is also nonsingular and diagonal, and may therefore be expressed in the form 
\begin{eqnarray} \label{}
\beta = U_1^{T}\alpha U_2,
\end{eqnarray}
where $U_1$ and $U_1$ are unitary matrices representing rotations of the local basis. Stage (3) of the proof is the concluded by considering the density matrices in each local subsystem,
\begin{subequations}
\begin{align}
&\rho^{(1)} = \beta\beta^{\dagger}, \label{} \\ 
&\rho^{(2)} = \beta^{\dagger}\beta, \label{}
\end{align}
\end{subequations}
which are therefore diagonal such that
\begin{eqnarray} \label{}
\rho^{(1)}_{ij} = \rho^{(2)}_{ij} = |\gamma_i|^2\delta_{ij}.
\end{eqnarray}
However, the set of characteristic values $\left\{|\gamma_i|\right\}$ of any matrix is basis independent and so may be identified with $\left\{|\alpha_i|\right\}$ in one of three ways;

\begin{enumerate}

\item If all $|\alpha_i|$ are distinct, there is a one-to-one correspondence between the set of eigenvalues $\left\{|\alpha_i|\right\}$ and the set of local eigenvectors $\left\{\Ket{i}_1\right\}$ (or $\left\{\Ket{i}_2\right\}$). Therefore $\Ket{\Psi}$ has canonical form for only one set of basis eigenstates. %$\left\{\Ket{i}_1,\Ket{i}_2\right\}$. 
In this case, all that a nondemolition experiment can verify is the basis in which $\Ket{\Psi}$ has canonical form.

\item Some $|\alpha_i|$ are equal. This allows us to verify the basis in which $\Ket{\Psi}$ has canonical form \emph{and} to specify the relative phases between different sets of $\alpha_i$ with equal magnitudes.

\item If \emph{all} $|a_i|$ are equal we may specify the phases of all $a_i$, find the value of $|\alpha_i|$ from normalisation and, consequently, specify the state completely.

\end{enumerate}

We have at last shown that \emph{only} nonlocal states which can brought into canonical form with equal coefficients (\ref{states}) may be verified using nondemolition measurements of nonlocal operators which are of the general kind developed by Aharonov, Albert and Vaidman. Assuming these to be the \emph{only} kinds of measurements that are possible without violating causality, as \emph{was} assumed by the authors themselves in 1985, it would be true to say that \emph{only} nonlocal operators of the form (\ref{class1}), (\ref{class2}) or (\ref{class3}) could be granted the status of observables in a relativistic theory, and that only states of the form (\ref{states}) could be verified by the their measurement.

However, sixteen years later Berry Groisman and Benni Reznik \cite{GroismanReznik2002}, also of Tel Aviv university, showed that the physical role of measurement in quantum mechanics need not be as restricted as the \emph{ideal measurements of the first kind} proposed by Von Neumann \cite{VonNeumann1955} and the early pioneers of the theory. They concluded that, if a measurement is not necessarily required to prepare the system in an eigenstate of the corresponding operator, a large number of observables, previously excluded on the grounds of causality violation, become measurable. This work, in turn, drew on the results of a paper published in 1994 by Lev Vaidman and Sandu Popescu of the University of Brussels \cite{PopescuVaidman1994}. The paper proved two important theorems which hold true for \emph{any} measurement of a nonlocal variable which is required to be consistent with relativistic causality. We will now briefly consider these two theorems.

%%%%%%%%%%%%%%%%%%%%%%%%%%%%%%%%%%%%%%%%%%%%
%%%%%%%%%%%%%%%%%%%%%%%%%%%%%%%%%%%%%%%%%%%%
\section{Further developments (and more questions)}
\label{SectVI}

%%%%%%%%%%%%%%%%%%%%%%%%%%%%%%%%%%%%%%%%%%%%%%%%%%%``
\subsection{This procedure is very specific, can we find general restrictions which apply to any measurement of a nonlocal variable? ``Yes" - Popescu and Vaidman (1994)}
\label{SectVI.1}

Popescu and Vaidman considered the following general scheme for a bipartite system,

\begin{itemize}

\item $t << t_0$, prepare the nonlocal state $\Ket{\Psi}$

\item $t = t_0 - \epsilon$, local interaction at subsystem $2$ described by the unitary transformation $\hat{U}^{(2)}$

\item $t = t_0$, perform a state verification measurement on $\Ket{\Psi}$

\item $t = t_0 + \epsilon$, perform a measurement of a local operator $\hat{A}^{(1)}$ on subsystem $1$.

\end{itemize}

They assumed as little as possible about the way in which the nonlocal measurement could be carried out. In particular they did not specify that the measurement should be nondemolition and, in fact, assumed nothing about the final state of the system following the measurement. Nor did they assume only local interactions between the system and the measuring device, as Aharonov and Albert had. Instead they considered only that the system should undergo unitary time evolution during the interaction. Their results therefore describe completely \emph{general} properties of the measurements of nonlocal variables.

The definition of state the verification measurement they adopted may be summarised as follows; if the measurement is to verify whether $\Ket{\Psi} = \Ket{\Psi_0}$, then the result of the experiment is ``Yes" if $\Ket{\Psi} = \Ket{\Psi_0}$ and ``No" if $\Ket{\Psi} = \Ket{\Psi_0^{\perp}}$, a state orthogonal to $ \Ket{\Psi_0}$. If, in general, the state of the system is described by 
\begin{eqnarray} \label{Psi}
\Ket{\Psi} = \alpha\Ket{\Psi_0}  + \beta\Ket{\Psi_0^{\perp}}. 
\end{eqnarray}
the result will be ``Yes" or ``No" with probabilities $|\alpha_1|^2$ and $|\alpha_2|^2$, respectively. The state $\Ket{\Psi_0}$ may then be written in canonical form 
\begin{eqnarray} \label{}
\Ket{\Psi_0} = \sum_{i}\alpha_{i}\Ket{i}_1  \otimes \Ket{i}_2, 
\end{eqnarray}
for some basis, where $\left\{\Ket{i}_1\right\}$ and $\left\{\Ket{i}_2\right\}$ span the subspaces $H_0^{(1)} \subset H^{(1)}$ and $H_0^{(2)} \subset H^{(2)}$.

In order to state the two theorems it is also convenient to rewrite the causality condition in a more compact form using the notation,
\begin{eqnarray} \label{}
p\left(\Psi\right) = Prob\left(\hat{A}^{(1)}=a, \ at \ t = t_0 + \epsilon \bigg| \Ket{\Psi} at \ t \ll t_0;  \ non-local \ measurement \ at \ t=t_0\right),
\end{eqnarray}
where $p\left(\Psi\right)$ represents the probability of getting the result $a$ for the measurement of $\hat{A}^{(1)}$ at $t = t_0 + \epsilon$, given that the state $\Ket{\Psi}$ was prepared initially at $ t \ll t_0$ and a nonlocal verification measurement was carried out at $ t=t_0$.  Causality then requires that the probability of getting the result $a$ must be independent of local interactions at subsystem $2$ according to the equation 
\begin{eqnarray} \label{Causal}
p\left(\hat{U}^{(2)}\Psi\right) = p\left(\Psi\right).
\end{eqnarray}
If the state of the system is described by (\ref{Psi}) and the initial state of the measuring device is given by $\Ket{\Phi}$, we may rewrite (\ref{Causal}) as
\begin{eqnarray} \label{Causal*}
\Bra{\Phi}\bra{\Psi_0^{\perp}}\hat{U}^{(2)\dagger}\hat{U}^{\dagger}\hat{P}^{(1)}_a\hat{U}\hat{U}^{(2)}\Ket{\Psi_0}\Ket{\Phi} = \Bra{\Phi}\bra{\Psi_0^{\perp}}\hat{U}^{\dagger}\hat{P}^{(1)}_a\hat{U}\Ket{\Psi_0}\Ket{\Phi},
\end{eqnarray}
where $\hat{U}$ describes the unitary evolution during the nonlocal verification and $\hat{P}^{(1)}_a$ is the standard projection operator onto the states with eigenvalue a in subsystem $1$. Theorems $1$ and $2$ may then be stated as follows.\\
\\
\indent
\emph{Theorem 1:} If $\Ket{\Psi} \in H^{(1)}_0 \otimes H^{(2)}$ then $p\left(\Psi\right)=p\left(\Psi_0\right)$, which is equal to a constant independent of $\Ket{\Psi}$. In general, however,
\begin{eqnarray} \label{Psi*}
\Ket{\Psi} = \alpha\Ket{\Psi'}  + \beta\Ket{\Psi''}, 
\end{eqnarray}
where $\Ket{\Psi'}$ and $\Ket{\Psi''}$ are the normalised projections onto $H^{(1)}_0 \otimes H^{(2)}$ and $\left(H^{(1)} - H^{(1)}_0\right) \otimes H^{(2)}$, respectively. The probabilities of local measurements on subsystem $1$ after the nonlocal measurement may therefore depend on $\Ket{\Psi}$, but only through $\beta\Ket{\Psi''}$.\\
\\
\indent
\emph{Theorem 2:} If $\Ket{\Psi} = \alpha\Ket{\Psi'}  + \beta\Ket{\Psi''}$, as in (\ref{Psi*}), then
\begin{eqnarray} \label{}
p\left(\Psi\right) = |\alpha|^2p\left(\Psi_0\right) + |\beta|^2p\left(\Psi''\right).
\end{eqnarray}
We now quickly summarise the proofs of these two theorems before discussing their wider implications in the context of our investigation.\\
\\
\indent
{\it Proof of Theorem 1:} The causality condition (\ref{Causal*}) combined with our definition of a reliable verification measurement above gives,
\begin{eqnarray} \label{EqualsZero}
\Bra{\Phi}\bra{\Psi_0^{\perp}}\hat{U}^{\dagger}\hat{P}^{(1)}_a\hat{U}\Ket{\Psi_0}\Ket{\Phi} = 0, \ \forall \Ket{\Phi}.
\end{eqnarray}
The set $\left\{\hat{U}\Ket{\Psi_0}\Ket{\Phi}\right\}$ therefore forms the subspace of states of the \emph{system + measuring
device} which will yield the answer ``Yes" in our state verification measurement. Conversely, $\left\{\hat{U}\Ket{\Psi_0^{\perp}}\Ket{\Phi}\right\}$ forms the subspace of ``No" states. Then, as $\hat{P}^{(1)}_a$ acts only on the system and not the measuring device, all states of the form $\hat{P}^{(1)}_a\hat{U}\Ket{\Psi_0}\Ket{\Phi}$ will also belong to the subspace of ``Yes" states, and are therefore perpendicular to states of the form $\hat{U}\Ket{\Psi_0^{\perp}}\Ket{\Phi}$. The proof of theorem $1$ may then be divided into one lemma and two propositions.\\
\\
\indent
{\it Lemma 1:} If $\Ket{\Psi} = \alpha\Ket{\Psi_0}  + \beta\Ket{\Psi_0^{\perp}}$ as in (\ref{Psi}) then $p\left(\Psi\right)=p\left(\Psi_0\right)$ if and only if $p\left(\Psi_0^{\perp}\right)=p\left(\Psi_0\right)$.\\
\\
\indent
{\it Proof:} Using (\ref{Psi}) and (\ref{Causal*}) 
%the form of $p\left(\Psi\right)$ in (6.4) and (6.1) above 
we obtain,
\begin{eqnarray} \label{long}
p\left(\Psi\right) = |\alpha|^2p\left(\Psi_0\right) + |\beta|^2p\left(\Psi_0^{\perp}\right) + \alpha\beta^{*}\Bra{\Phi}\bra{\Psi_0^{\perp}}\hat{U}^{\dagger}\hat{P}^{(1)}_a\hat{U}\Ket{\Psi_0}\Ket{\Phi} + \alpha^{*}\beta\Bra{\Phi}\Bra{\Psi_0}\hat{U}^{\dagger}\hat{P}^{(1)}_a\hat{U}\ket{\Psi_0^{\perp}}\Ket{\Phi}.
\end{eqnarray}
Using (\ref{EqualsZero}) together with the normalisation condition $|\alpha|^2 + |\beta|^2 = 1$, this reduces to
\begin{eqnarray} \label{}
p\left(\Psi\right) = |\alpha|^2 p\left(\Psi_0\right) + |\beta|^2\left(p\left(\Psi_0^{\perp}\right)-p\left(\Psi_0\right)\right),
\end{eqnarray}
from which we may see that $p\left(\Psi\right)=p\left(\Psi_0\right)$ if and only if $p\left(\Psi_0^{\perp}\right)=p\left(\Psi_0\right)$, as stated.\\
\\
\indent
{\it Proposition 1:} If $\Ket{\Psi} \in H^{(1)}_0 \otimes H^{(2)}$ and may therefore be expressed in the form 
\begin{eqnarray} \label{need}
\Ket{\Psi}  = \sum_{i=1}^{N}c_{i}\hat{U}^{(2)}_{i}\Ket{\Psi_0}, 
\end{eqnarray}
where $\Ket{\Psi_0} \in H^{(1)}_0 \otimes H^{(2)}_0$ as before, then
\begin{eqnarray} \label{ProbRel}
p\left(\sum_{i=1}^{N}c_{i}\hat{U}^{(2)}_{i}\Ket{\Psi_0}\right) = p\left(\Psi_0\right)
\end{eqnarray}
and the probabilities of local measurements on subsystem 1 are the same as if the initial state had been $\Ket{\Psi_0}$.\\
\\
\indent
{\it Proof:} Proposition $1$ may be proved by induction. If $N = 1$ and $c_1 = 1$ we recover the causality condition (\ref{Causal}). We may then assume that the relation (\ref{ProbRel}) is true for $N = n$ and consider the case for $N = n + 1$, where causality implies
\begin{eqnarray} \label{Equality}
p\left(\sum_{i=1}^{n+1}c_{i}\hat{U}^{(2)}_{i}\Ket{\Psi_0}\right) 
%&=& p\left((U^{(2)}_{n+1})^{-1}\sum_{i=1}^{n+1}c_{i}U^{(2)}_{i}\Ket{\Psi_0}\right)
%\nonumber\\ &=& 
= p\left((\hat{U}^{(2)}_{n+1})^{-1}\sum_{i=1}^{n+1}c_{i}\hat{U}^{(2)}_{i}\Ket{\Psi_0} + c_{n+1}\Ket{\Psi_0}\right).
\end{eqnarray}
Now, if we normalise the state $(\hat{U}^{(2)}_{n+1})^{-1}\sum_{i=1}^{n+1}c_{i}\hat{U}^{(2)}_{i}\Ket{\Psi_0}$ by introducing a normalisation factor $\mathcal{N}$ we obtain,
\begin{eqnarray} \label{Probs}
p\left(\mathcal{N}\sum_{i=1}^{n+1} c_{i}(\hat{U}^{(2)}_{n+1})^{-1}\Ket{\Psi_0}\right) = p\left(\Psi_0\right).
\end{eqnarray}
Identifying this state with the general state (\ref{Psi}) gives,
\begin{eqnarray} \label{}
\mathcal{N}\sum_{i=1}^{n+1} c_{i}(U^{(2)}_{n+1})^{-1}\Ket{\Psi_0} = \alpha\Ket{\Psi_0} + \beta\Ket{\Psi_0^{\perp}},
\end{eqnarray}
and, together with Lemma $1$, Eq. (\ref{Probs}) becomes $p\left(\Psi\right) = p\left(\Psi_0\right)$. Now, decomposing the term on the right-hand-side of Eq. (\ref{Equality}) such that
\begin{eqnarray} \label{}
(\hat{U}^{(2)}_{n+1})^{-1}\sum_{i=1}^{n+1}c_{i}\hat{U}^{(2)}_{i}\Ket{\Psi_0} + c_{n+1}\Ket{\Psi_0} = \left(\frac{\alpha}{\mathcal{N}} + c_{n+1}\right)\Ket{\Psi_0} + \frac{\beta}{\mathcal{N}}\Ket{\Psi_0^{\perp}},
\end{eqnarray}
and again identifying with (\ref{Psi}), giving $\Ket{\Psi} = \alpha'\Ket{\Psi} + \beta'\Ket{\Psi_0^{\perp}}$, where $\alpha' = (\alpha/\mathcal{N}) + c_{n+1}$ and $\beta' = (\beta/\mathcal{N})$, we see that
\begin{eqnarray} \label{}
p\left((\hat{U}^{(2)}_{n+1})^{-1}\sum_{i=1}^{n+1}c_{i}\hat{U}^{(2)}_{i}\Ket{\Psi_0} + c_{n+1}\Ket{\Psi_0}\right) = p\left(\Psi_0\right).
\end{eqnarray}
Therefore, in general, $p\left(\Psi\right) = p\left(\Psi_0\right)$ if $\Ket{\Psi} = \sum_{i=1}^{N}c_{i}\hat{U}^{(2)}_{i}\Ket{\Psi_0} \in H^{(1)}_0 \otimes H^{(2)}$.\\
\\
\indent
{\it Proposition 2:} To complete the proof of Theorem $1$ we must show that (\ref{ProbRel}) implies that $\Ket{\Psi} \in \Ket{\Psi_0} \in H^{(1)}_0 \otimes H^{(2)}$, which we have so far assumed. To do this it is sufficient to prove that superpositions of the form (\ref{ProbRel}) can express any vector belonging to the subspace $H^{(1)}_0 \otimes H^{(2)}$. This in turn may be done by showing that (\ref{ProbRel}) can express any one of a set of basis vectors spanning the subspace. By considering the Schmidt decomposition of $\Ket{\Psi}$ it is possible to show that a set of unitary transformations acting on the individual bases $\left\{\Ket{p}_1\right\}$ of $H^{(1)}$ and $\left\{\Ket{q}_2\right\}$ of $H^{(2)}$ may be defined such that superpositions of $\hat{U}_i^{(2)}\Ket{\Psi_0}$, as in (\ref{ProbRel}), yield vectors of the form $\Ket{p}_1 \otimes \Ket{q}_2 \in H^{(1)}_0 \otimes H^{(2)}$ which form such a basis. The demonstration is straightforward, but will omitted here for the sake of brevity and we refer the interested reader to \cite{PopescuVaidman1994}. We may therefore move on to the proof of Theorem $2$.\\
\\
\indent
{\it Proof of Theorem 2:} Now using (\ref{Psi*}) instead of (\ref{Psi}) and following a procedure analogous to that used to obtain equation (\ref{long}) above we see that,
\begin{eqnarray} \label{long*}
p\left(\Psi\right) = |\alpha|^2p\left(\Psi'\right) + |\beta|^2p\left(\Psi''\right) + \alpha\beta^{*}\Bra{\Phi}\Bra{\Psi''}\hat{U}^{\dagger}\hat{P}^{(1)}_a\hat{U}\Ket{\Psi'}\Ket{\Phi} + \alpha^{*}\beta\Bra{\Phi}\Bra{\Psi'}\hat{U}^{\dagger}\hat{P}^{(1)}_a\hat{U}\Ket{\Psi''}\Ket{\Phi}.
\end{eqnarray}
The last two terms of (\ref{long*}) are complex conjugates of each other and so to prove Theorem $2$ we need only show that one or other of them is equal to zero. Given that $\Ket{\Psi'} \in H^{(1)}_0 \otimes H^{(2)}$ we may substitute in from (\ref{need}), giving
\begin{eqnarray} \label{complex}
\alpha\beta^{*}\Bra{\Phi}\Bra{\Psi''}\hat{U}^{\dagger}\hat{P}^{(1)}_a\hat{U}\Ket{\Psi'}\Ket{\Phi} = \alpha\beta^{*}\sum_{i=1}^{N}c_i\Bra{\Phi}\Bra{\Psi''}\hat{U}^{\dagger}\hat{P}^{(1)}_a\hat{U}\hat{U}^{(2)}_{i}\Ket{\Psi'}\Ket{\Phi}
\end{eqnarray}
The proof is completed by showing that each term on the right hand side of (\ref{complex}) must be equal to zero by causality. Applying (\ref{Causal*}) to the terms in the right hand side of (\ref{complex})  we obtain 
\begin{eqnarray} \label{}
\Bra{\Phi}\Bra{\Psi''}\hat{U}^{\dagger}\hat{P}^{(1)}_a\hat{U}\hat{U}^{(2)}_i\Ket{\Psi_0}\Ket{\Phi} 
= \Bra{\Phi}\Bra{\Psi''}\hat{U}^{(2)}_i\hat{U}^{\dagger}\hat{P}^{(1)}_a\hat{U}\Ket{\Psi_0}\Ket{\Phi}. 
\end{eqnarray}
Now, we see that $\Ket{\Psi''} \in \left(H^{(1)}-H^{(1)}_0\right) \otimes H^{(2)}$ and so $(\hat{U}^{(2)}_i)^{-1}\Ket{\Psi''} \in \left(H^{(1)}-H^{(1)}_0\right) \otimes H^{(2)}$ as $(\hat{U}^{(2)}_i)^{-1}$ acts only on subsystem $2$. Thus $(\hat{U}^{(2)}_i)^{-1}\Ket{\Psi''}$ is orthogonal to $\Ket{\Psi_0}$ and each term in the right hand side of (\ref{complex}) vanishes.

%%%%%%%%%%%%%%%%%%%%%%%%%%%%%%%%%%%%%%%%%%%%%%%%%%%``
\subsection{Are there still nonlocal variables that can't be measured without violating causality? - ``Yes "}
\label{SectVI.2}

{\it Physical significance of Theorems 1 and 2:} The consequence of these theorems is that local information about the part of the initial state that lies in $H^{(1)}_0 \otimes H^{(2)}$ in necessarily erased. This is therefore an unavoidable feature of any verification measurement of a nonlocal state. 

This was thought to have profound implications for the measurability of operators in relativistic quantum mechanics. For example, the measurement of an operator $\hat{A}$ may be seen as a verification measurement of each of its nondegenerate eigenstates. It follows from Theorems 1 and 2 that the final state of the system after a measurement of $\hat{A}$ must be locally independent of its initial state. In addition, in accordance with the standard interpretation of the postulates of quantum mechanics, if the initial state is an eigenstate of $\hat{A}$  then it should remain undisturbed by the measurement.

This is an important point. Although Vaidman and Popescu did not assume that their state verification measurement necessarily left the state undisturbed, they considered that the measurability of an individual variable $A$, which may be seen here as equivalent to the measurability of a collection of state verifications, depended on the eigenstates of the operator $\hat{A}$ remaining invariant during the measurement process. Applying this assumption, in conjunction with Theorems 1 and 2, to the simplest nonlocal system, our two- fermion spin-state system, it is easy to  ``prove" that many operators are not causally measurable.

In fact, as we will later see, all operators which represent physical values (local or nonlocal) possessed by the system \emph{are} measurable. However, Theorems 1 and 2 restrict the way in which these variables may be measured. In particular, certain projective measurements \emph{do} violate causality, and are therefore deemed unphysical. For example, suppose $\hat{P}_{\Ket{\Psi_0}}$  is the projection onto the subspace of entangled states 
\begin{eqnarray} \label{Entangled}
\Ket{\Psi_0} = \alpha\Ket{\uparrow_{z}}_1 \otimes \Ket{\uparrow_{z'}}_2 + \beta\Ket{\downarrow_{z}}_1 \otimes \Ket{\downarrow_{z'}}_2 = \alpha\Ket{\Psi_1} + \beta\Ket{\Psi_2},
\end{eqnarray}
where these may represent an \emph{arbitrary} entangled state by appropriate choice of local basis.

Now, $\Ket{\Psi_1}$ and $\Ket{\Psi_2}$ are both eigenstates of $\hat{P}_{\Ket{\Psi_0}}$ corresponding to the eigenvalue zero and so are not disturbed by the measurement. Consequently the state will end in either $\Ket{\Psi_1}$ or $\Ket{\Psi_2}$ which are locally indistinguishable. However, according to Theorem 1, all local information is erased, which leads to a contradiction. The implication is then that  $\hat{P}_{\Ket{\Psi_0}}$ is unmeasurable. 

Similar arguments imply that only operators with nonlocal eigenstates which are maximally entangled are causally measurable. If a general entangled state (\ref{Entangled}) is an eigenstate of $\hat{A}$ then, by Theorem 1, the eigenstates of $\hat{A}$ must be locally indistinguishable, which is only the case if $|\alpha| = |\beta| = 1/\sqrt{2}$.

A more disturbing aspect of these assumptions is that even some operators with eigenstates which are local states, are apparently not causally measurable. In fact, so called \emph{ideal measurements} of operators whose eigenstates are combinations of direct products of spin states aligned along two or three independent axis, say $z$,$z'$ and $z"$, violate causality.

At first sight it may seem that we have therefore finally answered question (2) of Sect. \ref{SectIV.4} and, furthermore, that we have answered it in the negative. Aharonov and Albert claimed that only maximally entangled states were causally verifiable because they were the only nonlocal states verifiable using their original methods, and they did not, at that time, propose any others. Here it may seem that we have shown this result is true for \emph{any} verification procedure. But, in this last analysis, the additional assumption that operator measurements in quantum theory must be \emph{ideal measurements of the first kind} has crept in.

We may now see that, if the requirement that the measurement of an observable must prepare the system in an eigenstate of that observable is dropped, then a much wider class of nonlocal variables may be measured causally in accordance with Theorems 1 and 2, which are concerned with the erasing of local information.

%%%%%%%%%%%%%%%%%%%%%%%%%%%%%%%%%%%%%%%%%%%%%%%%%%%``
%%%%%%%%%%%%%%%%%%%%%%%%%%%%%%%%%%%%%%%%%%%%%%%%%%%``
\section{Can a larger class of nonlocal variables ever be measurable in a relativistic theory? - ``Yes, but only if we reconsider the role of measurement", Groisman and Reznik (2002)}
\label{SectVII}

Groisman and Reznik pointed out that quantum measurements play a dual role. The first is to allow us to observe the value of an unknown quantity, and the second is to prepare the system in a particular state. They argued that the roles of observation and preparation are, in fact, logically independent. 

The measurement technique they developed is based on the idea of \emph{remote operations} \cite{Eisert_etal2002} and does not assume that the measurement of an observable necessarily prepares the system an eigenstate of that same observable. They showed that, using this method, the class of causally measurable nonlocal operators could be greatly extended. In particular they showed that \emph{all} Hermitian operators for a $(2 \otimes 2)$-dimensional Hilbert space are in fact measurable.

The procedure follows the same general two stage protocol as the instantaneous measurements devised by Aharonov and Albert. However, in this method it is necessary for the observers share a \emph{large} supply of distributed entangled pairs. The general protocol is then as follows.\\
\\
\indent
\emph{Stage (1)}: Each observer, Alice or Bob, applies an interaction between his/her system and a set of ancillary particles. He/she then measures a set of local quantities whose values are recorded classically. This first part of the measurement is instantaneous as the interaction time $\Delta t \rightarrow 0$, even if the distance between subsystems $A$ and $B$ is $L \gg c\Delta t$.\\
\\
\indent
\emph{Stage (2)}: The results of the local measurements are combined via classical information exchange and the result of the nonlocal variable is known.\\
\\
\indent
Again, although the second stage requires a finite amount of time, the measurement itself is instantaneous as the value of the nonlocal variable is instantaneously ``encoded" in the correlations between the two space-like separated classical systems. Both entanglement and local operations are used here to produce a \emph{remote instantaneous transformation}  \cite{Eisert_etal2002} which maps a locally unmeasurable set of eigenstates to a locally measurable set. This is in contrast to Aharonov and Albert's procedure which uses entanglement and local operations to produce correlations between nonlocal states and locally measurable ones, but which does not \emph{map} one set of states to another. It therefore leaves the initial state intact whereas, in Groisman and Reznik's, procedure the initial state is necessarily destroyed.

The principle of causality combined with nonlocal action, as demonstrated in quantum entanglement, requires that the map is not deterministic \cite{RohrlichPopescu} and different mappings are generated with varying probabilities. However, in all cases it is possible
to infer the ``unmeasurable" nonlocal states from the locally measurable ones. The process itself is best illustrated by an example, and we will now demonstrate how it may be used to measure a nonlocal operator whose eigenstates are the $2 \otimes 2$ twisted product basis 
\begin{subequations}
\begin{align}
&\Ket{\Psi_1}_{AB} = \Ket{\uparrow_{z}}_A \otimes \Ket{\uparrow_{z}}_B, \label{1*} \\
&\Ket{\Psi_2}_{AB} = \Ket{\uparrow_{z}}_A \otimes \Ket{\downarrow_{z}}_B, \label{2*} \\
&\Ket{\Psi_3}_{AB} = \frac{1}{\sqrt{2}}\Ket{\downarrow_{z}}_A \otimes \left(\Ket{\uparrow_{z}}_B + \Ket{\downarrow_{z}}_B\right), \label{3*} 
\\
&\Ket{\Psi_3}_{AB} = \frac{1}{\sqrt{2}}\Ket{\downarrow_{z}}_A \otimes \left(\Ket{\uparrow_{z}}_B - \Ket{\downarrow_{z}}_B\right). \label{4*} 
\end{align}
\end{subequations}
The example is important as \emph{ideal} measurements of this operator, considered previously in Sect. \ref{SectVI.2}, were shown to violate causality \cite{PopescuVaidman1994}.\\
\\
\indent
{\it The process in detail:} Initially, Bob and Alice share one ancillary entangled pair, or \emph{ebit}, denoted with lower case letters $a$ and $b$. This gives the initial state,
\begin{eqnarray} \label{}
\Ket{\Psi} = \frac{1}{\sqrt{2}}\left(\Ket{\uparrow_{z}}_a \otimes \Ket{\uparrow_{z}}_b + \Ket{\downarrow_{z}}_a \otimes \Ket{\downarrow_{z}}_b\right) \otimes \Ket{\Psi}_{AB}.
\end{eqnarray}
\\
\indent
{\it Stage (1):} Bob performs a local C-NOT interaction with respect to the component of spin along the $y$-axis (here denoted by $\mathbb{I}_A \otimes \hat{\sigma}^{(B)}_y$) between the entangled qubit $b$ and his state $B$. This is described by the unitary transformation 
\begin{eqnarray} \label{}
\hat{U} = \Ket{\uparrow_{z}}_b \otimes \Ket{\uparrow_{z}}_b \otimes \mathbb{I}_B + \Ket{\downarrow_{z}}_b \otimes \Ket{\downarrow_{z}}_b \otimes \hat{\sigma}^{(B)}_y
\end{eqnarray}
and yields the state
\begin{eqnarray} \label{}
\Ket{\Psi} = \frac{1}{\sqrt{2}}\left(\Ket{\downarrow_{z}}_a \otimes \Ket{\downarrow_{z}}_b \otimes \mathbb{I}_B + \Ket{\uparrow_{z}}_a \otimes \Ket{\uparrow_{z}}_b \otimes \hat{\sigma}^{(B)}_y\right) \otimes \Ket{\Psi}_{AB}.
\end{eqnarray}
He then measures the $x$-component of spin of the entangled qubit $\hat{\sigma}^{(b)}_x$ (in the following notation some tensor products with the identity matrix with the Pauli spin matrices will be omitted) and records the result $\nu(\hat{\sigma}^{(b)}_x)$. The state is now described by
\begin{eqnarray} \label{}
\Ket{\Psi} = \left(\Ket{\downarrow_{z}}_a \otimes \mathbb{I}_B \pm \Ket{\uparrow_{z}}_a \otimes \hat{\sigma}^{(B)}_y\right) \otimes \Ket{\Psi}_{AB} = S \Ket{\Psi}_{AB},
\end{eqnarray}
where the $\pm$ corresponds to the two possible values of $\nu(\hat{\sigma}^{(b)}_x)$ and $S$ is called the \emph{state operator} or \emph{stator}, which satisfies the eigen-operator equation
\begin{eqnarray} \label{}
\hat{\sigma}^{(a)}_xS = \nu(\hat{\sigma}^{(a)}_x)\hat{\sigma}^{(B)}_y S.
\end{eqnarray}
This equation describes the correlations between unitary transformations performed by Alice on $a$ and the equivalent rotations on Bob's state. In particular, the transformation $\exp(i\alpha\hat{\sigma}^{(a)}_x)$ performed by Alice, is equivalent to a unitary transformation given by $\exp(i\alpha\hat{\sigma}^{(B)}_y)$ on Bob's qubit. Having prepared $S$ by the procedures above, Alice now measures $\hat{\sigma}^{(A)}_z$, with two possible outcomes,
\begin{subequations}
\begin{align}
&\Ket{\Psi}_A = \Ket{\downarrow_{z}}_a \label{Down}, \\ %\Ket{0}_A, 
&\Ket{\Psi}_A = \Ket{\uparrow_{z}}_a \label{Up}. %\Ket{1}_A.
\end{align}
\end{subequations}
If $\Ket{\Psi}_A =  \Ket{\downarrow_{z}}_A$ as in (\ref{Down}) she then measures $\hat{\sigma}^{(a)}_z$ and keeps the result $\nu(\hat{\sigma}^{(a)}_z)$. This induces the transformation 
\begin{eqnarray} \label{}
\left(\frac{1+\hat{\sigma}^{(a)}_z}{2}\right) \otimes \mathbb{I}_B + \hat{\sigma}^{(b)}_x\left(\frac{1-\nu(\hat{\sigma}^{(a)}_z)}{2}\right) \otimes \hat{\sigma}^{(B)}_y.
\end{eqnarray}
on Bob's qubit. If $\Ket{\Psi}_A =  \Ket{\uparrow_{z}}_a$ as in (\ref{Up}), she instead untwists Bob's cubit by performing a rotation of $\exp(i\pi\hat{\sigma}^{(a)}_x/4)$ before measuring $\hat{\sigma}^{(a)}_z$, as before, which induces the remote transformation 
\begin{eqnarray} \label{CondRot}
\left[\left(\frac{1+\hat{\sigma}^{(a)}_z}{2}\right) \otimes \mathbb{I}_B + \hat{\sigma}^{(b)}_x\left(\frac{1-\nu(\hat{\sigma}^{(a)}_z)}{2}\right) \otimes \hat{\sigma}^{(B)}_y\right] \exp\left(i\frac{\pi}{4}\nu(\hat{\sigma}^{(b)}_x)\hat{\sigma}^{(B)}_y\right).
\end{eqnarray}
This process is equivalent to a conditional $\pi/2$ rotation of Bob's state when Alice's state is $\Ket{\downarrow_{z}}_A$ which maps the twisted basis on Bob's side according to (\ref{CondRot}) (see also Eqs. (\ref{1*})-(\ref{4*})) with no rotation if Alice's state is $\Ket{\uparrow_{z}}_A$;
\begin{eqnarray} \label{CondRot*}
\left\{\Ket{\downarrow_{z}}_B + \Ket{\uparrow_{z}}_B,\Ket{\downarrow_{z}}_B - \Ket{\uparrow_{z}}_B\right\} \rightarrow \left\{\Ket{\downarrow_{z}}_B,\Ket{\uparrow_{z}}_B\right\}.
\end{eqnarray}
The four possible outcomes of this map are therefore those given in Table $1$ below.
\begin{center}
\begin{tabular}{|c|c|c|}
\hline
$\hat{\sigma}^{(a)}_z$/$\hat{\sigma}^{(b)}_x$ & $\nu(\hat{\sigma}^{(b)}_x) = +\hbar/2$ & $\nu(\hat{\sigma}^{(b)}_x) = -\hbar/2$\\
\hline
$\nu(\hat{\sigma}^{(a)}_z) = +\hbar/2$ &\vtop{\hbox{\strut $\Ket{\Psi_1}_{AB} \rightarrow \Ket{\downarrow_{z}}_A \otimes \Ket{\downarrow_{z}}_B$}\hbox{\strut $\Ket{\Psi_2}_{AB} \rightarrow \Ket{\downarrow_{z}}_A \otimes \Ket{\uparrow_{z}}_B$}\hbox{\strut $\Ket{\Psi_3}_{AB} \rightarrow \Ket{\uparrow_{z}}_A \otimes \Ket{\downarrow_{z}}_B$}\hbox{\strut $\Ket{\Psi_4}_{AB} \rightarrow \Ket{\uparrow_{z}}_A \otimes \Ket{\uparrow_{z}}_B$}}&\vtop{\hbox{\strut $\Ket{\Psi_1}_{AB} \rightarrow \Ket{\downarrow_{z}}_A \otimes \Ket{\downarrow_{z}}_B$}\hbox{\strut $\Ket{\Psi_2}_{AB} \rightarrow \Ket{\downarrow_{z}}_A \otimes \Ket{\uparrow_{z}}_B$}\hbox{\strut $\Ket{\Psi_3}_{AB} \rightarrow \Ket{\uparrow_{z}}_A \otimes \Ket{\uparrow_{z}}_B$}\hbox{\strut $\Ket{\Psi_4}_{AB} \rightarrow \Ket{\uparrow_{z}}_A \otimes \Ket{\downarrow_{z}}_B$}}\\
\hline
$\nu(\hat{\sigma}^{(a)}_z) = -\hbar/2$ &\vtop{\hbox{\strut $\Ket{\Psi_1}_{AB} \rightarrow \Ket{\downarrow_{z}}_A \otimes \Ket{\uparrow_{z}}_B$}\hbox{\strut $\Ket{\Psi_2}_{AB} \rightarrow \Ket{\downarrow_{z}}_A \otimes \Ket{\downarrow_{z}}_B$}\hbox{\strut $\Ket{\Psi_3}_{AB} \rightarrow \Ket{\uparrow_{z}}_A \otimes \Ket{\uparrow_{z}}_B$}\hbox{\strut $\Ket{\Psi_4}_{AB} \rightarrow \Ket{\uparrow_{z}}_A \otimes \Ket{\downarrow_{z}}_B$}}&\vtop{\hbox{\strut $\Ket{\Psi_1}_{AB} \rightarrow \Ket{\downarrow_{z}}_A \otimes \Ket{\uparrow_{z}}_B$}\hbox{\strut $\Ket{\Psi_2}_{AB} \rightarrow \Ket{\downarrow_{z}}_A \otimes \Ket{\downarrow_{z}}_B$}\hbox{\strut $\Ket{\Psi_3}_{AB} \rightarrow \Ket{\uparrow_{z}}_A \otimes \Ket{\downarrow_{z}}_B$}\hbox{\strut $\Ket{\Psi_4}_{AB} \rightarrow \Ket{\uparrow_{z}}_A \otimes \Ket{\uparrow_{z}}_B$}}\\
\hline
\end{tabular}
\end{center}
Finally, Bob measures the operator $\hat{\sigma}^{(B)}_z$ for his state $\Ket{\Psi}_B$.\\
\\
\indent
{\it Stage 2:} Alice and Bob then communicate their results to one another classically and use the values of $\nu(\hat{\sigma}^{(a)}_z)$ and $\nu(\hat{\sigma}^{(b)}_x)$ to identify which of the four mappings in Table $1$ has occurred. The values of $\nu(\hat{\sigma}^{(A)}_z)$  and $\nu(\hat{\sigma}^{(B)}_z)$ then allow them to infer the initial state of the system. In addition we see that all local information about the initial state of the system is necessarily erased, in accordance with Theorems $1$ and $2$ from Popescu and Vaidman \cite{PopescuVaidman1994}.
With appropriate modifications this procedure can also be used to measure operators whose eigenstates are the general $2 \otimes 2$ product basis. That is, when the coefficients $1/\sqrt{2}$ in equations (\ref{3*}) and (\ref{4*}) are replaced by $(\cos(\alpha/2),\sin(\alpha/2))$ and $(\sin(\alpha/2),\cos(\alpha/2))$, respectively. Here the procedure is necessarily more complicated as we succeed in obtaining the rotation (\ref{CondRot}) only with probability $1/2$. With probability $1/2$ we also obtain the alternative, less useful map 
\begin{eqnarray} \label{CondRot*A}
\left\{\Ket{\downarrow_{z}}_B + \Ket{\uparrow_{z}}_B,\Ket{\downarrow_{z}}_B - \Ket{\uparrow_{z}}_B\right\} \rightarrow \left\{\sin(\alpha/2)\Ket{\downarrow_{z}}_B +\cos(\alpha/2)\Ket{\uparrow_{z}}_B,\cos(\alpha/2)\Ket{\downarrow_{z}}_B - \sin(\alpha/2)\Ket{\uparrow_{z}}_B\right\}.
\end{eqnarray}
However, if Bob measures $\nu(\hat{\sigma}^{(b)}_x) = +\hbar/2$ he knows that the map (\ref{CondRot}) occurred and that the experiment was successful. On the other hand if he measures $\nu(\hat{\sigma}^{(b)}_x) = -\hbar/2$, he and Alice may utilise a further entangled pair to perform an additional rotation by angle $\alpha$, which again gives the ``correct" map (\ref{CondRot}) with probability $1/2$, and so on. The
total probability for success is then $1/2,\ 3/4, \ . \ .$, which converges to one as the number of trials goes to infinity. However, if the rotation  $\alpha$ is chosen such that  $\alpha = \pi k/2^n$, the $n^{th}$ step will always succeed.

Perhaps most importantly, Groisman and Reznik also showed that, by appropriate choices of local measurements and unitary operations, the remote transfer method allows any nonlocal operator of a $(2 \otimes 2)$-dimensional Hilbert space with eigenvectors which are \emph{arbitrary nonmaximally entangled states} to be measured. In effect this last result implies that \emph{any} arbitrary $(2 \otimes 2)$-dimensional operator may be measured instantaneously without violating causality, and may therefore be granted the status of an observable.

With this result in mind, it would perhaps seem strange if arbitrary operators of higher dimensional Hilbert spaces were unmeasurable. Later that year it was indeed shown that a measurement procedure involving \emph{partial} quantum teleportation permitted the instantaneous measurement of any nonlocal operator belonging to a Hilbert space of arbitrary dimension \cite{Vaidman2003}.

%%%%%%%%%%%%%%%%%%%%%%%%%%%%%%%%%%%%%%%%%%%%%%%%%%%%%%%%%%%%%%%%%%%%%%%%%%
%%%%%%%%%%%%%%%%%%%%%%%%%%%%%%%%%%%%%%%%%%%%%%%%%%%%%%%%%%%%%%%%%%%%%%%%%%
\section{But can all nonlocal variables be granted the status of observables? - ``Yes! As long as we measure them in the right way" - Vaidman (2003)}
\label{SectVIII}

The technique of quantum teleportation \cite{Bennet_etal1993} is well documented and we will not consider it in detail here. Instead we shall only consider how \emph{part} of the teleportation process may be utilised to perform instantaneous measurements of nonlocal variables. Quantum teleportation itself is not instantaneous, as it relies on classical information transfer during an intermediate stage of the teleportation process. Vaidman's great breakthrough was to realise that performing a \emph{partial teleportation}, where our second observer Bob transfers his state to Alice via a Bell measurement, but does not tell her the result, preserves the quantum states along each axis. Alice then carries out local measurements independently of Bob and combining the results via classical information transfer at the end of the process completes the measurement of the nonlocal variable.

As an example, consider the measurement of a nonlocal operator in our two particle system whose eigenstates are direct products of spin states aligned along different axis, i.e. the $2 \otimes 2$ twisted product basis (\ref{1*})-(\ref{4*}). Standard projective/Von Neumann measurements of these operators were found by Sandu and Popescu to contradict causality, although Groisman and Reznik's technique rendered them measurable (c.f. Sects. \ref{SectVI.2} and \ref{SectVII}). Here we show how to measure operators of this form with the aid of \emph{partial teleportation}.

In our two particle system it is necessary that Alice and Bob must share the maximally entangled two-particle state, or \emph{singlet} state $\Ket{\Psi_{-}}_{AB}$, and that their friend Collin allows them to utilise a third system $C$ which he has prepared in an arbitrary state $\Ket{\Psi}_{C}$. The teleportation procedure is based on the identity below, so that a measurement by Bob in the Bell basis collapses the joint $ABC$ wave function to one of the terms on the right hand side of the equation
\begin{eqnarray} \label{}
\Ket{\Psi}_{C}\Ket{\Psi_{-}}_{AB} = \frac{1}{2}\left(\Ket{\Psi_{-}}_{CA} \otimes \Ket{\Psi}_{B} + \ket{\Psi_{+}}_{CA} \otimes \ket{\tilde{\Psi}^{(z)}}_{B} + \Ket{\Phi_{-}}_{CA} \otimes \ket{\tilde{\Psi}^{(x)}}_{B} + \Ket{\Phi_{+}}_{CA} \otimes \ket{\tilde{\Psi}^{(y)}}_{B}\right),
\end{eqnarray}
where 
\begin{subequations}
\begin{align}
&\Ket{\Psi_{\pm}}_{ij}  = \frac{1}{\sqrt{2}}\left(\Ket{\uparrow}_{i} \otimes \Ket{\downarrow}_{j} \pm \Ket{\downarrow}_{i} \otimes \Ket{\uparrow}_{j} \right), 
\label{} \\
&\Ket{\Phi_{\pm}}_{ij}  = \frac{1}{\sqrt{2}}\left(\Ket{\uparrow}_{i} \otimes \Ket{\uparrow}_{j} \pm \Ket{\downarrow}_{i} \otimes \Ket{\downarrow}_{j} \right),
\end{align}
\end{subequations}
and $\ket{\tilde{\Psi}^{(k)}}$ denotes a rotation of $\Ket{\Psi}$ by $\pi$ about the $k^{th}$ axis, thereby effectively transferring Bob's state to Alice.

The key step in measuring a nonlocal variable is then a local measurement of $\hat{\sigma}^{(A)}_z$ by Alice, which she may carry out at any time, independently of Bob. If the result is $\hat{\sigma}^{(A)}_z = +\hbar/2$, this is equivalent to measuring $\hat{\sigma}^{(B)}_z$ at Bob's site, and if the result is $\hat{\sigma}^{(A)}_z = -\hbar/2$ it is equivalent to measuring $\hat{\sigma}^{(B)}_x$. Combining the results of their local measurements Alice and Bob (and Collin) are then able to distinguish unambiguously between the four eigenstates (\ref{1*})-(\ref{4*}). With appropriate modifications it is, in principle possible, to use this technique to measure all nonlocal operators acting on a $(2\otimes2)$-dimensional Hilbert space, which confirms Groisman and Reznik's conclusion that all such operators may be granted the status of observables in a relativistic quantum theory.

However, the most important aspect of this technique is that it may also be extended to measure \emph{any} arbitrary function $\hat{O}(\hat{q}_A,\hat{q}_B, . \ . \ . \hat{q}_N)$ of $N$ local observables $\left\{\hat{q}\right\}$. That is, in principle, it may be used to measure any nonlocal variable of a composite quantum system, even if the subsystems $A,B, \ . \ . \ . \ N$ are space-like separated and $N$ is arbitrarily large. Again, local information about the initial state of the system is erased in accordance with Popescu and Vaidman's results \cite{PopescuVaidman1994}.

The general protocol for the measurement of an arbitrary nonlocal operator in the two particle system $\hat{O}(\hat{q}_A,\hat{q}_B)$ was set out in detail by Vaidman in his 2003 paper \cite{Vaidman2003}. The process is complicated and consists of four individual steps, each performed in a number of teleportation ``rounds" (from now on we will use the term teleportation to mean partial teleportation in the sense explained above). Each successive round requires the utilisation of increasing numbers of teleportation channels, that is of increasing numbers of ``Collins" and shared entangled pairs. However, by considering the two-particle case it is relatively easy to see how the procedure may be continued for three, four, five . . . subsystems, and it is therefore illustrative for us to consider it in detail.

Firstly, Alice and Bob swap the states of their systems with the states of $K$ spin-$1/2$ particles using teleportation. The protocol is then as follows.\\
\\
\indent
{\it Round 1}

\begin{enumerate}

\item Bob teleports the state of his system (which is equivalent to the states of $K$ spin-$1/2$ particles) to Alice and records the outcomes of the associated Bell measurements. The number of possible outcomes is $N = 4^K$ which Bob indexes by the variable $n = 1,2, \ . \ . \ N$, where $n = 1$ corresponds to measuring singlets in all Bell measurements and indicates that Bob's state has been successfully teleported without distortion.

\item Alice performs a unitary operation on the composite system of herself and the teleported spins. If undistorted teleportation has occurred this operation transforms the eigenstates of $\hat{O}(\hat{q}_A,\hat{q}_B)$, which are now all at Alice's site, to product states of the $z$-component of spin.

\item Alice teleports the complete system of $2K$ spin-$1/2$ particles to Bob.

\item If undistorted teleportation has occurred $(n = 1)$, Bob now measures the teleported system in the $z$-basis and his local spin measurement in the $z$-basis completes the measurement of $\hat{O}(\hat{q}_A,\hat{q}_B)$.

However, the probability of obtaining singlet states in all Bell measurements, that is of obtaining teleportation without distortion, is $1/N$ and so the measurement will only ``succeed" on the fourth step one in $N$ times.

\end{enumerate}

{\it Round 2}

\begin{enumerate}

\item If $n \neq 1$, Bob teleports the system back to Alice and again records the outcomes of the associated Bell measurements, which he indexes with the variable $m_1 = 1,2, \ . \ . \ M$ where $M = 4^{2K}$. In this case Bob must also ``tell" Alice the outcomes of his previous Bell measurements $n$ via the same teleportation channel.

\item Alice can then perform unitary operations which consist of her original transformations from the first round of teleportations, together with corrections required to correct the distortion, on each system in a further $N -1$ teleportation channels which she will use in a second round. Again, if no distortion has occurred $(m_1=1)$ Alice's operations transform the eigenstates of $\hat{O}(\hat{q}_A,\hat{q}_B)$ into product $z$-spin states.

\item Alice teleports all $N -1$ systems back to Bob.

\item If no distortion has occurred $(m_1 = 1)$, Bob measures the teleported system in the $z$-spin basis and the measurement of $\hat{O}(\hat{q}_A,\hat{q}_B)$ is complete. This occurs with probability $1/M$.

\end{enumerate}

{\it Round 3}

\begin{enumerate}

\item If $m_1 \neq 1$, Bob again teleports his system back to Alice in the teleportation channel corresponding to the ``subcluster" $m_1$ of the original cluster $n$, together with his results $m_1$, and records the outcomes of the latest set of Bell measurements $m_2$.

\item Alice performs unitary operations on each system in $(N -1)(M -1)$ teleportation channels used in the third round. The operations on each channel are such that if Bob teleported the system in this channel and the teleportation succeeded without distortion, the eigenstates of $\hat{O}(\hat{q}_A,\hat{q}_B)$ are again transformed to $z$-spin product states.

\item Alice teleports all $(N -1)(M -1)$ systems back to Bob.

\item If $m_2 = 1$, Bob measures the teleported system in the $z$-basis and the measurement is complete. This occurs with probability $1/M$ as in the second round.

\end{enumerate}

Again, if $m_2 \neq 1$, Bob teleports the system back to Alice . . . , and so on, until a successful measurement is performed. Thus we see that the general procedure requires many teleportation channels which are utilised sequentially in a series of nested clusters, with two channels in the first round, $4^K -1$ clusters of two channels in the second round, $4^{2K} -1$ subclusters of two channels in the third . . . etc. The probability for success in the first round is $1/4^K$, and is $1/4^{2K}$ in every subsequent round, so that the total probability for success may be brought arbitrarily close to one if sufficient entanglement resources are available.

The procedure may be extended to three parts in the following manner. In step $1$ Bob and Collin, who is now allowed to take part in the experiments and not to just assist, both teleport their states to Alice. In stage $2$ Alice completes unitary operations which require undistorted teleportation from both Bob and Collin in order to transform the eigenstates of an operator $\hat{O}(\hat{q}_A,\hat{q}_B,\hat{q}_C)$ into $z$-spin product states. In step $3$ she teleports the complete system to Bob, as before. In step $4$, instead of performing local measurements, Bob teleports the system to Collin using a particular teleportation channel $n_B$, which is dependent upon the outcome of his first set of Bell measurement, $n$ .

There are then five additional stages. In step $5$ Collin teleports all the systems from Bob's teleportation channels, except the system corresponding to $(n_B, n_c) = (1,1)$, back to Alice. Instead he measures this system in the $z$-spin basis (step $6$) and chooses a particular channel $(n_B, n_c)$ to send the original system from channel $n_B$, according to the results of his Bell measurements $n_C$. In the final three stages Alice again performs appropriate unitary operations (step $7$) and, assuming no distortion teleports them back to Bob (step $8$), who performs local measurements (step $9$) and starts again at stage $1$ to initiate the second round if necessary. An analogous procedure involving $4 \times 4 = 16$ individual steps may be used to measure a nonlocal variable of a four part system, $5 \times 5 = 25$ steps for a five part system, and so on.

Finally we notice that Vaidman's procedure in no way contradicts the principle of relativistic causality, and that the question posed way back in 1931 has finally has finally been answered - \emph{all} nonlocal variables may measured instantaneously be granted the status of observables in relativistic quantum mechanics! \emph{Or can they . . . ?}

\section{Discussion}
\label{SectIX}

\subsection{Conclusions}
\label{SectIX.1}

We have seen that there are, in principle, \emph{no} causal restrictions upon the measurability of self-adjoint operators in quantum theory. In particular the values of all nonlocal variables may be determined instantaneously, and so all such variables may be granted the status of quantum mechanical observables. This, in turn, implies that all nonlocal states may be verified by specifying their simultaneous eigenvalues for sets of commuting operators.

However, we have also seen that relativistic causality \emph{does} place restrictions upon the way in which nonlocal, and even some local variables may be measured. It was found that the standard Von Neumann interpretation of quantum mechanical measurement was too restrictive, and that certain projective measurements which would allow superluminal signaling are not physically realisable (c.f. Sect. \ref{SectVI.2}).

In such cases other forms of measurement have been discovered, which do not violate causality, but which do not necessarily prepare the system in an eigenstate of the measured observable (Sects. \ref{SectVII}-\ref{SectVIII}). Similarly, state verification measurements performed using these procedures do not preserve the initial state, but may only ascertain its simultaneous eigenvalues. Correspondingly, some nonlocal states, may not be verified in nondemolition experiments (Sects. \ref{SectIV.3}, \ref{SectV}). It was also shown that any measurement of a nonlocal variable necessarily erases \emph{some} local information, which is a completely general result (Sect. \ref{SectVI.1}).

\subsection{Questions for the future}
\label{SectIX.2}

These conclusions hold, for the time being, with one important caveat. It is, in fact, not yet clear whether \emph{all} variables related to spread out fermionic wave functions may be measured instantaneously. In particular, it is not clear whether the above results can be generalised to include all quantum systems which are themselves in a superposition of being in different places. Measurements of variables related to such systems rely on the ability to transform the local state of a particle, which is in a nonlocal superposition, into a locally measurable state of a composite system; that is, of the system consisting of the particle(s) and measuring device. Although, for bosons, local operations of this form have already been achieved \cite{Bennet_etal1993}, the space-like separated local variables corresponding to the measurement of a fermion state should, in theory, fulfill anticommutation relations in accordance with the Pauli exclusion principle. This has led Vaidman and others to postulate the existence of superselection rules which prevent such transformations \cite{Maitre1997}.

%\begin{trivlist} 
%\end{trivlist}

\section{Acknowledgements}

This essay was originally submitted in partial fulfillment of the requirements of the Certificate of Advanced Study in Mathematics (aka ``Part III")  examination at the University of Cambridge, in the academic year 2005-2006. The essay supervisor was Dr. Berry Groisman who, as well as suggesting the title and basic content of the essay, generously gave his time to explain some of the finer, more intriguing, and more perplexing aspects of quantum theory to a hapless Part III student. It was, and is, much appreciated.

%\appendix
%\section{}
%\label{}

%\addcontentsline{toc}{section}{References}
%\addcontentsline{toc}{chapter}{Bibliography}

%%%%%%%%%%%%%%%%%%%%%%%%%%%%%%%%%%%%%%%%%%%%%
\section{Bibliography}

\begin{enumerate}

\item A. I. M. Rae, {\it Quantum Mechanics}, $4^{th}$ ed, Institute of Physics Publishing (2002).

\item C. J. Isham, {\it Lectures on Quantum Theory, Mathematical and Structural Foundations}, Imperial College Press (1995).

\item P.A.M. Dirac, {\it Principles of Quantum Mechanics}, $4^{th}$ ed, International Series of Monographs on Physics, Oxford University Press (1958).

\item P.A.M. Dirac, {\it Lectures on Quantum Mechanics}, Dover Publications (2001).

\item J. Von Neumann, {\it Mathematical Foundations of Quantum Mechanics}, Princeton Landmarks in Mathematics and Physics, Princeton University Press (1955).

\item 
J. Preskill {\it Lecture Notes for Physics 229: Quantum Information and Computation}, \\ http://www2.fiit.stuba.sk/$\sim$kvasnicka/QuantumComputing/ %PreskilTextbook_all.pdf 
(1998); \\
J. Preskill, {\it Lecture Notes for Physics 219: Quantum Computation}, \\ http://www.lorentz.leidenuniv.nl/quantumcomputers/literature/
%preskill_9.pdf 
(2004).

\item M. A Nielsen and I. L. Chuang, {\it Quantum Computation and Quantum Information}, $1^{st}$ ed., Cambridge University Press (2000).

\item Y. Aharonov and D. Rohrlich, {\it Quantum paradoxes: Quantum Theory for the Perplexed}, WILEY-VCH Verlag (2005). %GmbH and Co (2005).

\end{enumerate}

\end{document}